\documentclass[12pt,preprint]{aastex}

\begin{document}

\title{Jet Properties of GeV-Selected Radio-Loud Narrow-Line Seyfert 1 Galaxies and Possible Connection to Their Disk and Corona}
\author{Xiao-Na Sun\altaffilmark{1,3}, Jin Zhang\altaffilmark{2,3}, Da-Bin Lin\altaffilmark{1}, Zi-Wei Xue\altaffilmark{2}, En-Wei Liang\altaffilmark{1,2,3, 4}, Shuang-Nan Zhang\altaffilmark{2,4,5}}
\altaffiltext{1}{Guangxi Key Laboratory for Relativistic Astronomy, Department of Physics, Guangxi University, Nanning 530004, China}
\altaffiltext{2}{National Astronomical Observatories, Chinese Academy of Sciences, Beijing 100012, China; zhang.jin@hotmail.com}
\altaffiltext{3}{Key Laboratory for the Structure and Evolution of Celestial Objects, Chinese Academy of Sciences, Kunming 650011, China}\altaffiltext{4}{Key Laboratory of Particle
Astrophysics, Institute of High Energy Physics, Chinese Academy of Sciences, Beijing 100049,
China}\altaffiltext{5}{Physics Department, University of Alabama in
Huntsville, Huntsville, AL 35899, USA}

\begin{abstract}
The observed spectral energy distributions of five GeV-selected narrow-line Seyfert 1 (NLS1) galaxies are fitted with a model including the radiation ingredients from the relativistic jet, the accretion disk, and the corona. We compare the properties of these GeV NLS1 galaxies with flat spectrum radio quasars (FSRQs), BL Lacertae objects (BL Lacs) and radio-quiet (RQ) Seyfert galaxies, and explore possible hints for jet-disk/corona connection. Our results show that the radiation physics and the jet properties of the GeV NLS1 galaxies resemble that of FSRQs. The luminosity variations of PMN J0948+0022 and 1H 0323+342 at GeV band is tightly correlated with the beaming factor ($\delta$), similar to that observed in FSRQ 3C 279. The accretion disk luminosities and the jet powers of the GeV NLS1 galaxies cover both the ranges of FSRQs and BL Lacs. With detection of bright corona emission in 1H 0323+342, we show that the ratio of the corona luminosity ($L_{\rm corona}$) to the accretion disk luminosity ($L_{\rm d}$) is marginally within the high end of this ratio distribution for a RQ Seyfert galaxy sample, and the variation of jet luminosity may connect with $L_{\rm corona}$. However, it is still unclear whether a system with a high $L_{\rm corona}/L_{\rm d}$ ratio prefers to power a jet.
\end{abstract}

\section{Introduction}
\label{sec:intro}
Narrow-line Seyfert 1 (NLS1) galaxies are a subclass of active galactic nuclei (AGNs). They are characterized by the narrow permitted lines (FWHM (H$\beta$) $<$ 2000 km s$^{-1}$), [O III]$\lambda$5007$/$H$\beta$ $<$ 3, and the Fe II bump in their optical spectra (e.g., Boroson \& Green 1992; Pogge 2000). The masses of their central black holes (BHs) are systematically lower than those of broad-line Seyfert 1 (BLS1) galaxies (e.g., Boroson 2002; Komossa \& Xu 2007). The accretion of NLS1 galaxies may be close to the Eddington rate, or even super-Eddington rate (Boroson 2002; Collin \& Kawaguchi 2004; Xu et al. 2012). NLS1 galaxies are usually radio-quiet (RQ) and only three radio-loud (RL) NLS1 galaxies had been reported in literature before 2000 (Zhou et al. 2003). Systematic searches for RL NLS1 galaxies indicate that only 7\% of NLS1 galaxies are RL (Komossa et al. 2006; Zhou et al. 2006) and a very small fraction of them ($\sim2.5\%$) are `very' RL (radio-loudness $R_{\rm RL} >$ 100, Komossa 2008). Although the BH masses of NLS1 galaxies are usually lower than those in typical blazars by 1$-$2 orders of magnitude, they may be underestimated (Collin et al. 2006; Zhu et al. 2009; Calderone et al. 2013) and the BH masses of some RL NLS1 galaxies may be comparable to those in blazars (Zhou et al. 2003; Paliya et al. 2013a; Calderone et al. 2013). These RL NLS1 galaxies exhibit compact core-jet structure, flat/inverted radio spectra, and high brightness temperature (Komossa et al. 2006; Doi et al. 2006). It was proposed that most of the RL NLS1 galaxies display blazar characteristics and may also host relativistic jets (Zhou et al. 2003; Yuan et al. 2008).

Observations of the Large Area Telescope (LAT) on board the \emph{Fermi} satellite have confirmed that NLS1 galaxies are a new class of gamma-ray AGNs. Five NLS1 galaxies have been detected so far with {\em Fermi}/LAT (Abdo et al. 2009a; Calderone et al. 2011; D'Ammando et al. 2012) since the first detection of PMN J0948+0022 (Abdo et al. 2009b). Kiloparsec (kpc) scale radio structure is found in six NLS1 galaxies, including two GeV gamma-ray emitting NLS1 galaxies (PMN J0948+0022 and 1H 0323+342, Doi et al. 2012). Most interestingly, an apparent superluminal velocity of ($8.2\pm1.5$)$c$ in the jet of SBS 0846+513 was inferred with Very Long Baseline Array (VLBA) observations from 2011 to 2012 (D'Ammando et al. 2012). Those observations confirm the existence of relativistic jets in these RL NLS1 galaxies.

The properties and radiation mechanism of the jets in GeV RL NLS1 galaxies are hot topics in recent years. It is well known that the emission of blazars, including flat spectrum radio quasars (FSRQs) and BL Lacertae objects (BL Lacs), is dominated by their jet emission. Their broadband SEDs exhibit two distinct peaks, which can be well explained with the synchrotron radiation and the inverse Compton (IC) scattering of the relativistic electrons in the jets (Maraschi et al. 1992; Ghisellini et al. 1996; Zhang et al. 2012; 2014; Cao \& Wang 2013; Chen 2014; Kang et al. 2014; Yan et al. 2014). For FSRQs, the IC process of the photons in their broad-line regions (IC/BLRs) may dominate the IC peaks of their SEDs (e.g., Sikora et al. 1994; Ghisellini et al. 2009; 2010; Zhang et al. 2014). Some authors have reported that the properties of the broadband SEDs for GeV NLS1 galaxies are similar to blazars (Abdo et al. 2009a; Zhang et al. 2013a; Sun et al. 2013; Paliya et al. 2013a). By comparing the SEDs between GeV NLS1 galaxies PKS 1502+036 and PKS 2004$-$447 with blazars Mrk 421 and 3C 454.3, Paliya et al. (2013a) suggested that the SED properties of both PKS 1502+036 and PKS 2004$-$447 are similar to blazars and intermediate between FSRQs and BL Lacs. However, the steep $\gamma$-ray spectra of GeV NLS1 galaxies are more analogous to FSRQs than BL Lacs (Abdo et al. 2009a; Sun et al. 2013; Paliya et al. 2013a). The GeV RL NLS1 galaxies also display significant flux variations in the GeV gamma-ray band (e.g., Abdo et al. 2009b; Foschini et al. 2011; Paliya et al. 2013a). It is unclear what may be responsible for the flux variation and whether the GeV RL NLS1 galaxies have similar variation characters as those GeV$-$TeV blazars, in which the flux variation is always accompanied with the spectral shift (Zhang et al. 2012; 2013b).

The emission of RQ NLS1 galaxies is usually dominated by the emission from their accretion disks and coronas (e.g., Wang et al. 2004). The radiations of GeV NLS1 galaxies may consist of the radiations from their accretion disks and coronas as well as their jets. Therefore, the GeV NLS1 galaxies could be good probes to study the connections for jet emission and disk emission, hence the jet composition and formation mechanisms.

This paper investigates the radiation mechanism and the jet properties of GeV NLS1 galaxies. By collecting the broadband SED data from literature (Section 2), we fit the SEDs with the single-zone leptonic model plus the accretion disk and corona radiations using the minimum $\chi^2$ technique as described in Zhang et al. (2012, 2014) (Section 3). The derived jet parameters, including the Doppler factor, magnetic field strength, power, radiation efficiency, and magnetization parameter, and a comparison of jet properties for GeV RL NLS1 galaxies and blazars are shown in Section 4. The study of the possible connection between the relativistic jet and accretion disk is presented in Section 5. A comparison of corona emission for RL and RQ NLS1 galaxies is shown in Section 6. The conclusions and discussion are given in Section 7.

\section{Sample and Data}
\begin{figure*}
\includegraphics[angle=0,scale=0.55]{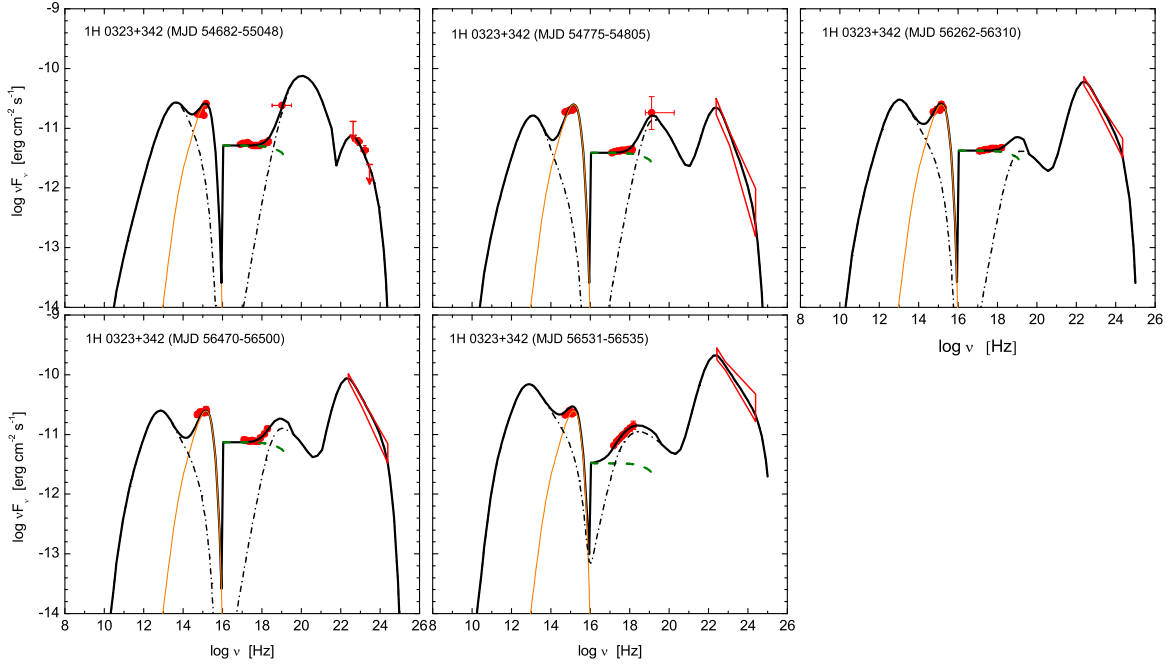}
\caption{\textbf{(a) Observed SEDs (red data points and bow-ties) with the model best-fits (lines) for 1H 0323+342 observed in five campaigns. The black thick solid lines are the sum of emission from the jet (dash-dotted lines), the accretion disk (orange thin solid lines) and the corona (magenta dashed lines). The steep rises in the black thick solid lines at $10^{16}$ Hz are due to the assumption of corona emission from $10^{16}$ Hz to 150 keV.}}\label{SEDs}
\end{figure*}

The five confirmed GeV RL NLS1 galaxies with \emph{Fermi}/LAT detection are included in our sample. We compile their simultaneously or quasi-simultaneously observed broadband SEDs form literature. They are shown in Figure \ref{SEDs}. The observations for these sources are summarized as below.

\emph{1H 0323+342} ($z=0.0629$, Zhou et al. 2007). Its optical spectrum is characterized with FWHM (H$\beta$)$=1600$ km s$^{-1}$, [OIII]/H$\beta\sim$0.12,
and a strong Fe II emission (Zhou et al. 2007). The source also presents remarkable hybrid behaviors of NLS1 galaxies and blazars (Zhou et al. 2007).
It is well known that the flat radio spectrum above 1 GHz and the compact core plus the one-sided jet structure (Beasley et al. 2002) with high
variabilities at the radio, optical, and X-ray bands, are typical features of blazars. 1H 0323+342 has a high radio loudness of $R_{\rm RL}=245$
(Doi et al. 2012). The optical observation with the \emph{Hubble Space Telescope} suggests that this source is hosted in a one-armed spiral galaxy
(Zhou et al. 2007), but Ant{\'o}n et al. (2008) reported that the host galaxy of this source has a ring structure triggered by an interacting/merging
process, which is consistent with the result of a 2D Fourier analysis of the imagery in Le{\'o}n
Tavares et al. (2014). With the very long baseline interferometer observations, Wajima et al. (2014) reported that the brightness temperature of this source is greater than $(5.2\pm0.3)\times10^{10}$ K and the derived equipartition Doppler factor is greater than 1.7 and the variability Doppler factor is 2.2. Its five broadband SEDs with {\em Fermi}/LAT observations in different campaigns are compiled from literature, as shown in Figure \ref{SEDs}(a). The observation data are taken from Abdo et al. (2009a) and Paliya et al. (2014).

\emph{SBS 0846+513} ($z = 0.5835$, Zhou et al. 2005). It was classified as a typical NLS1 galaxy by Zhou et al. (2005), who reported
FWHM (H$\beta$)=(1710$\pm$184) km s$^{-1}$, [OIII]/H$\beta\simeq$0.32, and a strong Fe II bump for this source with the Sloan Digital Sky Survey (SDSS)
data. This NLS1 galaxy was not detected at GeV band during the first two-year operation of \emph{Fermi}/LAT. A significant increase of the GeV flux was observed from 2010 October to 2011 August, and a strong flare at GeV band was detected by \emph{Fermi}/LAT in 2011 June (D'Ammando et al. 2012). Moreover, an apparent superluminal velocity of $(8.2\pm1.5)c$ in the jet was inferred with the VLBA images from 2011 to 2012 (D'Ammando et al. 2012), confirming the existence of a highly relativistic jet in this source. The broadband SED for this source is taken from D'Ammando et al. (2012), as shown in Figure \ref{SEDs}(b).

\emph{PMN J0948+0022} ($z=0.5846$, Abdo et al. 2009b). It is the first confirmed GeV RL NLS1 galaxy by \emph{Fermi}/LAT (Abdo et al. 2009b). Both the
radio power and radio loudness of this source fall in the range of the classical radio quasars with $P_{\rm rad} > 10^{26}$ W Hz$^{-1}$ and
$R_{\rm RL} > 1000$ (Zhou et al. 2003). Zhou et al. (2003) proposed that a relativistic jet may be hosted in the center of this object, which is
similar to blazars. A multiwavelength campaign from March to July in 2009 indicates that a flux decrease from optical to $\gamma$-ray bands was followed by an increase of radio emission (Abdo et al. 2009b), showing flux variation from radio to $\gamma$-ray bands. The broadband SEDs in different campaigns of this source were reported in Abdo et al. (2009b) and Foschini et al. (2012). We selected only those SEDs that have the confirmed {\em Fermi}/LAT detections from Abdo et al. (2009b) and Foschini et al. (2012). Eleven SEDs are selected for our analysis, as shown in Figure 1(c). Note that the SED observed on 2009 Aril 15 is consistent with that observed on 2009 June 14 within the observational errors, indicating that the source is roughly in the same stage during the two observational campaigns, as shown in Figure \ref{SEDs}(c). Similarly, the SEDs observed on 2009 May 25 and 2009 June 4 are also consistent within the error bars. Therefore, we obtain nine SEDs in different states.

\emph{PKS 1502+036} ($z=0.409$, Orienti et al. 2012). It is also a very RL NLS1 galaxy, with $R_{\rm RL}=1549$ (Yuan et al. 2008). The VLBA observations
at 15 GHz show that it has a core jet structure (Orienti et al. 2012). The intranight optical variability (Paliya et al. 2013b) and the rapid IR
variability (Jiang et al. 2012) were also detected. We get only one broadband SED from literature for this source, as shown in Figure \ref{SEDs}(b), which is taken from Paliya et al. (2013a).

\emph{PKS 2004$-$447} ($z=0.24$, Oshlack et al. 2001). It was identified as a candidate of NLS1 galaxy since its optical spectrum is consistent with
the criteria of NLS1 galaxy classification, i.e., FWHM (H$\beta$)$=1447$ km s$^{-1}$ and [OIII]/H$\beta\sim$1.6 (Drinkwater et al. 1997). The weak Fe II
lines make it classified as a narrow-line radio galaxy (Komossa et al. 2006). It was included in the CRATES catalog of flat spectrum objects
(Healey et al. 2007) due to its flat spectrum ($\alpha \sim 0.3$, $S(\nu) \propto \nu^{-\alpha}$) below 4.8 GHz, but it has a steep spectrum
($\alpha > 0.5$) above 8.4 GHz (Orienti et al. 2012). Observations show that the optical flux of PKS 2004$-$447 drops by a factor of $\sim$4 over an interval of several years and there is no evidence of a big blue bump (Oshlack et al. 2001). Abdo et al. (2009a) suggested that the jet power of PKS 2004$-$447 is well above the typical range of radio galaxies, favoring the hypothesis of a genuine NLS1 galaxy. We get only one broadband SED from literature for this source, as shown in Figure \ref{SEDs}(b). It is taken from Paliya et al. (2013a).

\section{SED Modeling and Results}
As mentioned in Section 1, the emission of the GeV NLS1 galaxies may be from the jet, the accretion disk, and the corona. We describe our model in details as follows.

\subsection{Jet Radiation}
The single-zone synchrotron+IC model is used to explain the jet emission, where the IC process includes both the synchrotron self-Compton (SSC) scattering and the external Compton (EC) scattering. The EC process may be dominated by the IC scattering of the BLR photons (IC/BLR) (Abdo et al. 2009a), similar to that in FSRQs (Ghisellini et al. 2009, 2010; Zhang et al. 2014). Thus the jet radiation is assumed to be dominated by the synchrotron radiation and the SSC process as well as the IC/BLR process of the relativistic electrons. We also assume that the relativistic jets of GeV NLS1 galaxies are close to the line of sight. Hence, the beaming factor ($\delta$) of the jet is taken as its bulk Lorentz factor ($\Gamma$), i.e., $\delta=\Gamma$. Note that the contributions of the IC processes from the tours, the accretion disk and the corona are not taken into account in our analysis. Some authors also considered the contribution of IC/disk in SED modeling for 1H 0323+342 (Abdo et al. 2009a; Paliya et al. 2014), and some authors only considered the contribution of IC/tours in SED modeling for SBS 0846+513 (D'Ammando et al. 2012), PKS 1502+036 (Paliya et al. 2013a) and PKS 2004$-$447 (Paliya et al. 2013a). However, the energy density of torus photon field is lower than that of BLR and the thermal emission of torus may also be much lower than the synchrotron radiation of the jet (Abdo et al. 2009a; Paliya et al. 2013a). In addition, the photons from the accretion disk are strongly redshifted (Sahayanathan \& Godambe 2012). Therefore, our model does not consider the thermal radiation of tours at infrared band and the IC processes from the tours, the accretion disk and the corona at GeV band.

The distribution of relativistic electrons is assumed as a broken power law (Ghisellini et al. 2009, 2010; Abdo et al. 2009a; Chen et al. 2012; Paliya et al. 2013a), which is characterized by an electron density parameter ($N_0$), a break energy $\gamma_{\rm b}$ and indices ($p_1$ and $p_2$) in the range of  $\gamma_{\rm e}$ to $[\gamma_{\min}, \gamma_{\max}]$. The radiation region is assumed to be a sphere with radius $R$, which is obtained with $R=\delta c\Delta t/(1+z)$, where $\Delta t$ is the variability timescale. For SBS 0846+513, we take $\Delta t=27.8$ hr (D'Ammando et al. 2012). The variability timescales of the other four objects are taken as $\Delta t=24$ hr (Paliya et al. 2013a).

In order to calculate the energy density of BLR photon field ($U_{\rm BLR}$) in the comoving frame, one needs to estimate its radius ($R_{\rm BLR}$) and total luminosity ($L_{\rm BLR}$). We estimate $L_{\rm BLR}$ using the luminosities of emission lines with Equation (1) given in Celotti et al. (1997). The size of the BLR is estimated using the BLR luminosity with Equation (23) in Liu \& Bai (2006). The fluxes of emission lines (H$\beta$ \& O III) for PKS 1502+036 and SBS 0846+513 are taken from Yuan et al. (2008), and the fluxes of the emission lines (Mg II, H$\gamma$ \& H$\beta$) for PMN J0948+0022 are selected from Zhou et al. (2003). Since the BLR is not uniform and has an angle to the disk, a corrected factor of 17/12 for $U_{\rm BLR}$ is also proposed (Ghisellini \& Madau 1996). The derived energy densities of the BLR photon fields in the comoving frame are $U_{\rm BLR}^{'}=2.48\times10^{-2}\Gamma^{2}$ erg cm$^{-2}$ for PKS 1502+036, $U_{\rm BLR}^{'}=1.90\times10^{-2}\Gamma^{2}$ erg cm$^{-2}$ for SBS 0846+513, and $U_{\rm BLR}^{'}=9.76\times10^{-3}\Gamma^{2}$ erg cm$^{-3}$ for PMN J0948+0022. Since no fluxes of emission lines are available in the literature for PKS 2004$-$447 and 1H 0323+342, we use the average energy density of the above three sources, and the corresponding value is $U_{\rm BLR}^{'}=1.79\times10^{-2}\Gamma^{2}$ erg cm$^{-2}$.

\subsection{Accretion Disk Radiation}
The accretion disk emission usually shows as a blue bump at $\sim 10^{15}$ Hz, which is modeled with the black body spectrum (e.g., Davis \& Laor 2011). This component is clearly detected in PKS 1502+036, 1H 0323+342, and PMN J0948+0022. We use the standard accretion disk spectrum (Davis \& Laor 2011) to explain this component. The parameters include the inside and outside radii ($R_{\rm in}$ and $R_{\rm out}$) of the accretion disk, BH mass, Eddington ratio, and inclination to the line of sight $i$. We take $R_{\rm out}=700R_{\rm s}$ ($R_{\rm s}$ is the Schwarzschild radius) and $\cos i=1$, and then vary both $R_{\rm in}$ (from 0.5$R_{\rm s}$ to 4.5$R_{\rm s}$) and the Eddington ratio to model the accretion disk emission in the ultraviolet band.

The BH mass is fixed in SED modeling. The estimate of the BH mass is quite uncertain. Using the data of H$\beta$ emission line, it was reported that $M_{\rm BH}\sim 1.8 \times10^7 M_{\odot}$ for 1H 0323+342 (Vestergaard \& Peterson 2006), $M_{\rm BH}\sim 8.2\times10^6 M_{\odot}$ for SBS 0846+513 (Zhou et al. 2005), $M_{\rm BH} \sim 4.0\times10^7 M_{\odot}$ for PMN J0948+0022 (Zhou et al. 2003), $M_{\rm BH}\sim4\times10^6 M_{\odot}$ for PKS 1502+036 (Yuan et al. 2008), and $M_{\rm BH}\sim5.4\times10^{6}M_{\odot}$ for PKS 2004$-$447 (Oshlack et al. 2001). It was suggested that the BH masses of NLS1 galaxies derived from the data of H$\beta$ emission line may be underestimated (Collin et al. 2006; Zhu et al. 2009; Calderone et al. 2013). For example, the BH mass of PMN J0948+0022 estimated with the H$\beta$ emission line and monochrome luminosity at 5100 ${\rm \AA}$ is $M_{\rm BH} \sim 4.0\times10^7 M_{\odot}$, but it is $\sim 8.1\times10^8 M_{\odot}$ with the Mg $_{\rm II}$ $\lambda$2798 and monochrome luminosity at 3000 ${\rm \AA}$ (Zhou et al. 2003). The BH mass of PKS 1502+036 estimated with data of the H$\beta$ emission line is $M_{\rm BH}\sim4\times10^6 M_{\odot}$ (Yuan et al. 2008), but it is $\sim 2.0\times10^8 M_{\odot}$ derived from the fit of big blue bump spectrum with the standard accretion disk model (Calderone et al. 2013). The BH masses of NLS1 galaxies may be comparable to that of blazars (e.g., Laor 2000; McLure \& Dunlop 2001; Calderone et al. 2013). Therefore, we model the accretion disk emission with $M_{\rm BH} \sim 8.1\times10^8 M_{\odot}$ (Zhou et al. 2003) for PMN J0948+0022 and $M_{\rm BH} \sim 2.0\times10^8 M_{\odot}$ (Calderone et al. 2013) for PKS 1502+036. For 1H 0323+342, we take $M_{\rm BH} \sim 4.0\times10^8 M_{\odot}$, which is the average value of BH mass for a sample of NLS1 galaxies reported in Calderone et al. (2013).

\subsection{Corona Radiation}
A flat spectrum component at the X-ray band is observed in 1H 0323+342. It is difficult to be explained with the emission from the jet and accretion disk. It may be attributed to the corona radiation (Abdo et al. 2009a). Therefore, a corona radiation model proposed by Ghisellini et al. (2009) is adopted here. The corona is assumed to be homogeneous between 3$R_{\rm s}$ and 30$R_{\rm s}$, and the radiation spectrum of corona is a cut-off power law, i.e., $L_{\rm X}(\nu) \propto \nu^{-\alpha_{\rm X}}\exp(-\nu/\nu_{\rm c})$, where the spectral index is fixed to $\alpha_{\rm X}=1$ and the high-energy cut-off frequency is taken as $\nu_{\rm c}=150$ keV.

\subsection{Fitting Strategy and Results}
The $\chi^2$ minimization technique is used to perform the SED fits as reported in Zhang et al. (2012, 2014). This technique requires the observed
errors of the SED data. Since no errors are available for some data points in the synchrotron radiation and SSC bumps of the selected SEDs, we take $10\%$ of the observation flux as the errors (Zhang et al. 2012) in order to obtain the goodness of the SED fit with the $\chi^2$ minimization technique. The upper limits of the data in SEDs are also used to constrain the model parameters, and the model prediction should be under the limits. For the details of this technique please refer to Zhang et al. (2012, 2014). The results of our SED fits are shown in Figure 1 and the derived model parameters with $1\sigma$ confidence level are reported in Table 1.

In our previous works (Zhang et al. 2013a, 2013c), we simply studied the jet emission component in the observed SEDs with the single-zone leptonic model for four GeV NLS1 galaxies (except for SBS 0846+513), and reported that the jet emission properties of GeV NLS1 galaxies are similar to FSRQs. In this paper we take the radiations of accretion disk and corona into account besides the jet emission, and the observation data and some model parameters (such as the variability timescale and the energy density of BLR) are not exactly the same as the previous works; therefore, one can find some differences for the derived jet parameters from the previous works. The observed broadband SEDs of GeV NLS1 galaxies have also been modeled with the single-zone leptonic models by other authors. We compare the derived physical parameters with those reported in the literature for these SEDs as follows: (1) For 1H 0323+342 in  Abdo et al. (2009a) and Paliya et al. (2014), besides the IC/BLR process, the contribution of IC/disk (and IC/torus) was considered. The derived $L_{\rm d}$, $B$ and $\Gamma$ are higher than ours, but $\gamma_{\min}$ and $\gamma_{\rm b}$ are smaller than ours; hence the peak frequency and luminosity of SSC bump in the model-predicted SED are lower than ours; (2) For SBS 0846+513 in D'Ammando et al. (2012), they only considered the IC/torus contribution in the EC process and thus obtained larger $\Gamma=15$ than us; (3) For PMN J0948+0022 in Abdo et al. (2009b) and Foschini et al. (2012), the model of Ghisellini \& Tavecchio (2009) was used; the derived parameters are roughly consistent with ours; (4) For PKS 1502+036 and PKS 2004$-$447 in Paliya et al. (2013a), the seed photons of IC process may be from the torus with an energy density of $\sim1\times10^{-3}$ erg cm$^{-3}$ in the AGN frame. Therefore, the derived $\Gamma$ and $\gamma_{\rm b}$ are slightly larger than our results, but $B$ is smaller.

\section{Comparison of Jet Properties for GeV NLS1 galaxies and Blazars}
With the data reported in Table 1, we show the distributions of $\delta$, $B$, $\gamma_{\rm b}$, and $\gamma_{\min}$ in comparison with BL Lacs and FSRQs in Figure \ref{jet-parameters}. The data of BL Lacs and FSRQs are from Zhang et al. (2014). They are also derived with the SED fitting by the single-zone leptonic models with the same technique. It is found that the $\delta$ distribution of GeV NLS1 galaxies ranges from 2.8 to 13.7 with a median of $\sim 9.5$, which is smaller than that in FSRQs and BL Lacs (with a median of $\sim$15). The distribution of the magnetic field strength ranges from 1.7 G to 5.8 G, with a median of $\sim 3.7$ G, which is systematically larger than that in the BL Lacs, but roughly close to that in FSRQs. The $\gamma_{\min}$ and $\gamma_{\rm b}$ distributions are also consistent with those in FSRQs. In the $\delta$$-$$\gamma_{\rm b}$ plane, both FSRQs and NLS1 galaxies occupy the same region, and a tentative anticorrelation is found for both of them, as illustrated in Figure \ref{gammab-delta}. The best linear fit in log scale gives $\gamma_{\rm b}\propto\delta^{-0.75\pm0.21}$. These results indicate that the jet properties of GeV NLS1 galaxies are analogous to those in FSRQs (see also Abdo et al. 2009a; Zhang et al. 2013c; Paliya et al. 2013a).

To further compare the jet properties for GeV NLS1 galaxies and blazars, we derive the jet powers by assuming that they are carried by relativistic electrons, cold protons, magnetic fields, and radiation (e.g., Ghisellini et al. 2009, 2010; Zhang et al. 2012, 2014), i.e., $P_{\rm jet}=\sum_i P_{\rm i}$, where $P_{\rm i}=\pi R^2 \Gamma^2 c U^{'}_{\rm i}$ and $U^{'}_{i} (i={\rm e,\ p,}\ B, \rm r)$ are the powers and the energy densities associated with the emitting electrons, cold protons, magnetic fields, and radiation (e.g., Zhang et al. 2014). The radiation efficiency and the magnetization parameter of a jet are calculated with $\varepsilon_{\rm r}=P_{\rm r}/P_{\rm jet}$ and $\sigma=P_{\rm B}/(P_{\rm p}+P_{\rm e}+P_{\rm r})$, respectively (Zhang et al. 2013c; 2014). The results are also reported in Table 2. The distributions of $P_{\rm jet}$, $\varepsilon_{\rm r}$ and $\sigma$ are shown in Figure \ref{Pjet}. One can observe that the $P_{\rm jet}$ distribution of GeV NLS1 galaxies roughly covers the intermediate region of FSRQs and BL Lacs, and the distributions of $\varepsilon_{\rm r}$ and $\sigma$ are roughly consistent with that of FSRQs. The medians of the $\varepsilon_{\rm r}$ and $\sigma$ distributions of GeV NLS1 galaxies are 0.13 and 0.42, respectively, and they are 0.22 and 0.59 for FSRQs. These results further confirm that the jet properties of GeV NLS1 galaxies are similar to those in FSRQs.

Violent variability is a characteristic of blazars. Similar feature is also observed in GeV NLS1 galaxies. For example, SBS 0846+513 was not detected by \emph{Fermi}/LAT during the first two-year operation and a strong flare at $\gamma$-ray band was observed in 2011 June (D'Ammando et al. 2012), indicating that this source has strong variability at GeV band. Calderone et al. (2011) analyzed the $\gamma$-ray lightcurves of the other four GeV NLS1 galaxies and reported variability timescales in the range of 3$-$30 days. They also suggested a hint for photon index variation, indicating that the flux variations are accompanied with the spectral variations similar to blazars (e.g., Zhang et al. 2013b). Zhang et al. (2013b) showed a tentative correlation between the gamma-ray luminosity and the Doppler factor for 3C 279. We examine whether the GeV NLS1 galaxies share the same relation with 3C 279 by using the observed SEDs of PMN J0948+0022 and 1H 0323+342 in different stages. As shown in Figure \ref{Lc-delta}, the peak luminosities of the EC bumps ($L_{\rm c}$) are tightly correlated with $\delta$ for the two sources. The $L_c-\delta$ relation observed in 3C 279 is also shown for comparison. The best linear fits in the log scale give $L_{\rm c}\propto\delta^{7.65\pm2.74}$ with a Pearson correlation coefficient of $r=0.91$ and chance probability of $p=7.7\times10^{-4}$ for PMN J0948+0022 and $L_{\rm c}\propto\delta^{4.12\pm1.52}$ with $r=0.98$ and $p=0.004$ for 1H 0323+342, respectively. The slopes of this relation in the three sources are roughly consistent, although the uncertainties of the slopes are very large with current data. This relation would be, at least partially, due to the Doppler boosting effect since the energy density of the external photon field is magnified by $\Gamma^2$ ($\sim\delta^2$) in the EC process (e.g., Zhang et al. 2013b).

\section{Connection Between Relativistic Jet and Accretion Disk}
The derived accretion disk luminosities of GeV NLS1 galaxies are $10^{44}-10^{46}$ erg s$^{-1}$. They are roughly intermediate between those of FSRQs and the upper limits of BL Lacs in comparison with a sample of blazars from Ghisellini et al. (2010). Note that the accretion disk emission bump does not show up in the SEDs of PKS 2004$-$447 and SBS 0846+513. Taking the luminosity at $10^{15}$ Hz of the model fitting line as the upper limits of their accretion disk luminosities, we get the upper limits of $10^{43.6}$ erg s$^{-1}$ for PKS 2004$-$447 and $10^{44.5}$ erg s$^{-1}$ for SBS 0846+513, respectively. Therefore, the true accretion disk luminosity distribution of GeV NLS1 galaxies may be in a more broad range than that of FSRQs, and even extends to the range of BL Lacs. The accretion disk luminosity in the Eddington units ($L_{\rm d}/L_{\rm Edd}$) of blazars is usually in the range from  $10^{-6}$ to 1 (Ghisellini et al. 2010), and Ghisellini et al. (2010) proposed a division of $L_{\rm d}/L_{\rm Edd}\sim0.01$ to separate BL Lacs and FSRQs. With the BH masses that are used to model the accretion disk emission in section 3.2, we get $L_{\rm d}/L_{\rm Edd}=0.01\sim 0.16$ for the GeV NLS1 galaxies, which is intermediate between FSRQs and BL Lacs. The upper limits of $L_{\rm d}/L_{\rm Edd}$ for PKS 2004$-$447 and SBS 0846+513 are $\sim 0.05$. With the division proposed by Ghisellini et al. (2010), the GeV NLS1 galaxies are analogous to FSRQs.

To reveal the connection between the jet power and the accretion disk luminosity, we show the GeV NLS1 galaxies in the $L_{\rm d}$$-$$P_{\rm jet}$ and $L_{\rm d}$$-$$P_{\rm r}$ planes in Figure \ref{Pjet-Ld} in comparison with a blazar sample taken from Ghisellini et al. (2010). One can find that the GeV NLS1 galaxies are in a broad range, scattering in the ranges of both FSRQs and BL Lacs. $P_{\rm jet}$ is higher than $L_{\rm d}$ for most of blazars in Ghisellini et al. (2010), but it is usually comparable and even lower than $L_{\rm d}$ for the GeV NLS1 galaxies. The GeV NLS1 galaxies tend to have a higher ratio of the disk luminosity to the jet power than blazars. From the $L_{\rm d}$$-$$P_{\rm r}$ plane, one can observe that the disk luminosities of the GeV NLS1 galaxies are higher than $P_{\rm r}$. The $L_{\rm d}$ of three GeV NLS1 galaxies whose accretion disk radiation bumps are clearly detected in their SEDs are 1$-$2 orders of magnitude higher than their $P_{\rm r}$, similar to FSRQs. We should also note that the upper limits of $L_{\rm d}$ for PKS 2004$-$447 and SBS 0846+513 indicate that their $L_{\rm d}$ are intrinsically dim, analogous to BL Lacs. Therefore, the GeV NLS1 galaxies are indeed a bridge between the FSRQs and BL Lacs.

\section{Comparison of Corona Emission for RL and RQ NLS1 galaxies}

Corona emission is clearly detected in 1H 0323+342. A power law spectrum with exponential cut-off proposed by Ghisellini et al. (2009) is adopted to fit this emission component. The derived corona luminosities are $2\sim 5\times 10^{44}$ erg s$^{-1}$, and the ratio of $L_{\rm corona}/L_{\rm d}$ is $0.45\sim 1.0$. The corona luminosities for a RQ Seyfert galaxy sample were reported in Wang et al. (2004). The corona luminosities of this sample distribute in $10^{41}$$-$$10^{45}$ erg s$^{-1}$ with a median of $6\times 10^{43}$ erg s$^{-1}$ and the $L_{\rm corona}/L_{\rm d}$ ratios range in 0.002$-$0.62 with a median of 0.036. Both $L_{\rm corona}$ and $L_{\rm corona}/L_{\rm d}$ of 1H 0323+342 are marginally in the high end of these distributions, as shown in Figure \ref{Lcorona}(a). The accretion disk emission component was clearly detected for PKS 1502+036 and PMN J0948+0022. We estimate their upper limits of $L_{\rm corona}$ with a ratio of $L_{\rm corona}/L_{\rm d}<$0.58, which is the median of $L_{\rm corona}/L_{\rm d}$ in different states of 1H 0323+342. They are also shown in the $L_{\rm corona}$$-$$L_{\rm d}$ plane \footnote{Since no significant accretion disk emission is observed in PKS 2004$-$447 and SBS 0846+513, we do not derive the upper limits of their corona emission and present in Figure \ref{Lcorona}.}. We also compare the GeV NLS1 galaxies with the RQ Seyfert galaxies in the $L_{\rm corona}/L_{\rm bol}-L_{\rm bol}/L_{\rm Edd}$ plane in Figure \ref{Lcorona}(b), where the bolometric luminosity ($L_{\rm bol}$) of RQ Seyfert galaxies is a proxy of their $L_{\rm d}$, but $L_{\rm bol}$ of the GeV NLS1 galaxies includes the radiations from both the accretion disk and the jet, i.e., $L_{\rm bol}=L_{\rm d}+L_{\rm jet}$. The GeV RL NLS1 galaxies are roughly consistent with the RQ Seyfert galaxies, and not distinguished from the RQ Seyfert galaxies as a unique population. These results likely imply that the disk-corona system in the RL Seyfert galaxies resembles to that in the RQ Seyfert galaxies.

\section{Conclusions and Discussion}
The observed SEDs of five GeV NLS1 galaxies are well explained with a model including the radiation ingredients from the relativistic jet, the accretion disk, and the corona. The single-zone leptonic model is used to fit the jet emission and the physical parameters of the jets, including $B$, $\delta$, $\gamma_{\min}$, $\gamma_b$, $P_{\rm jet}$, $\sigma$ and $\varepsilon_{\rm r}$, are derived from the fitting results. We compare their jet properties with those of BL Lacs and FSRQs. The distributions of these jet parameters indicate that the jet properties of the GeV NLS1 galaxies are intermediate between FSRQ jets and BL Lac jets, but more analogous to FSRQ jets. The derived accretion disk luminosities of 1H 0323+342, PKS 1502+036, and PMN J0948+0022 are $10^{44}$$-$$10^{46}$ erg s$^{-1}$, which are lower than those in FSRQs. On the other hand, the upper limits of $L_{\rm d}$ for PKS 2004$-$447 and SBS 0846+513 indicate that the radiation of accretion disk for some NLS1 galaxies may be intrinsically weak and similar to those in BL Lacs. The NLS1 galaxies are intermediate between FSRQs and BL Lacs in the $L_{\rm d}$$-$$P_{\rm jet}$ and $L_{\rm d}$$-$$P_{\rm r}$ planes. The corona emission was detected only in 1H 0323+342. The median of its $L_{\rm corona}/L_{\rm d}$ ratio is 0.58, which is marginally at the high end of the distribution of this ratio for a RQ Seyfert galaxy sample, implying that the disk-corona system in the RL Seyfert galaxies may resemble to that in the RQ Seyfert galaxies.

Significant flux variation is a feature of FSRQs and BL Lacs. This feature is also observed in the GeV NLS1 galaxies. As shown in Figure \ref{Lc-delta}, the peak luminosities of EC bumps at GeV band in PMN J0948+0022 and 1H 0323+342 are tightly correlated with $\delta$, similar to that observed in a typical FSRQ 3C 279 (Zhang et al. 2013b). This is likely due to the fact that the energy density of the external photon field is magnified by the Doppler factor in the EC process. Note that the accretion disk luminosities of PMN J0948+0022 and 1H 0323+342 in different states with different GeV luminosities keep almost constant, indicating that the luminosity variations of PMN J0948+0022 and 1H 0323+342 at the GeV band would not be due to the instability of the accretion disk. Using the observation data with \emph{BeppoSAX}, Grandi \& Palumbo (2004) untangled the radiations from the jet, accretion disk, and corona for 3C 273. They found that the emission from the corona becomes weak as the jet emission in the X-ray band increases and proposed a possible anti-correlation between the jet emission and corona emission. With the the detections of bright corona emission in 1H 0323+342, we examine whether it also shows similar feature. In order to compare with that observed in 3C 273 (Grandi \& Palumbo 2004), we derive the corona luminosity ($L_{\rm corona}$) in the energy band from $10^{16}$ Hz to 150 keV and the jet luminosity in the same energy band based on the best-fit results of the SED fits without considering the uncertainties of the model parameters. Figure \ref{Ljet-Lcorona}(a) shows $L_{\rm corona}$ as a function of $L_{\rm jet}$ in comparison with 3C 273. A similar feature as that shown in 3C 273 is found for 1H 0323+342, except for the data point obtained during the campaign MJD 56262-56310. Note that during the observation campaign of MJD 56262-56310, the emission at X-ray band is dominated by the corona radiation, and the jet emission in this energy band has great uncertainty. To further investigate this issue, we also show the beaming factor $\delta$ as a function of $L_{\rm corona}$ in Figure \ref{Ljet-Lcorona}(b). A weak anti-correlation between the two parameters is observed. These results imply that the instability of the corona may result in the variation of the physical condition of jets and lead to the variation of the jet emission. This also agrees with the discussion for the observed spectral variation in 3C 279 (Zhang et al. 2013b).

It has long been speculated that the physics in different BH jet systems may be essentially the same (e.g., Mirabel 2004; Zhang 2007). As shown in Zhang et al. (2013c), the correlation between the jet power and the prompt gamma-ray luminosity of gamma-ray bursts is consistent with the correlation between the jet power and the synchrotron radiation peak luminosity of FSRQs, and the GeV NLS1 galaxies extend this correlation to the low jet power end (see Figure 2 in Zhang et al. 2013c). The results in this paper further confirm that the radiation physics in these relativistic jets are analogous. The jets of GeV NLS1 galaxies are an intermediate group between FSRQs and BL Lacs. Note that the IC/BLR process is essential for the electron radiation cooling in both FSRQs and GeV NLS1 galaxies. The analogue of the emission properties for FSRQs and GeV NLS1 galaxies would be mainly due to this effect.

The formation mechanism of a relativistic jet in the BH system is a great issue. It is generally believed that the jet launching is connected with the accretion disk and corona in a source (Armitage \& Natarajan 1999; Merloni \& Fabian 2002; Cao 2004, 2014). The GeV NLS1 galaxies would be the best candidates to study this issue since the radiations from the jet, the accretion disk, and the corona may all be detected for these sources. With a very small sample in this analysis, we find that the accretion disk luminosity and the jet power of GeV NLS1 galaxies distribute in broad ranges, which may cover the ranges of both FSRQs and BL Lacs. It seems that there is not an universal jet formation mechanism at work among these GeV NLS1 galaxies, different from that in FSRQs, but may be similar to that in BL Lacs (Zhang et al. 2014). The significant corona emission is only detected in 1H 0323+342 among five GeV NLS1 galaxies. Although the $L_{\rm corona}/L_{\rm d}$ ratio of 1H 0323+342 distributes at the high end of the distribution of this ratio for a RQ Seyfert galaxy sample, it is still unclear whether a system with a high $L_{\rm corona}/L_{\rm d}$ ratio prefers to power a jet.

As an intermediate group between BL Lacs and FSRQs, GeV-selected NLS1 galaxies may shed light on the BL Lac$-$NLS1$-$FSRQ sequence. In analogy to the different states of BH X-ray binary, Zhang (2013) proposed a unified framework among different types of AGNs. Accretion rate is the fundamental of this framework, as shown in Figure 40 of Zhang (2013). In this framework, BL Lacs are characterized with a low accretion rate (with an Eddington ratio lower than $0.01$; Ghisellini et al. 2010; Xiong \& Zhang 2014). Their accretion disks may be truncated by the evaporation and are dominated by an clumpy Advection Dominated Accretion Flow (ADAF; e.g., Yuan \& Narayan 2014), and a small inner accretion disk that is fed by the condensation of corona gas may be presented (Liu et al. 2007; Liu 2013; Qiao \& Liu 2013; Qiao et al. 2013). As the accretion rate increases, with an Eddington ratio being $0.01\sim0.1$, the contribution of the inner disk increases. The accretion-outflow structure may transfer to that in GeV NLS1 galaxies. The inner thin disks for both of BL Lacs and GeV NLS1 galaxies may extend to around the inner-most stable circular orbit (ISCO), and thus their jets may be produced via the Blandford-Payne (BP, Blandford \& Payne 1982) and/or Blandford-Znajek (BZ, Blandford \& Znajek 1977) mechanisms (e.g., Zhang 2013; Zhang et al. 2014). The increase of the accretion rate may also lead to that the evaporation is not strong enough to completely remove the disk, then the corona is condensed by the inverse Compton scattering of the soft photons from the disk (Liu et al. 2007; Liu 2013). In this case, a complete thin disk with a weak corona would exist. The thin disk also extends to around ISCO and a jet may be launched mostly via the BZ mechanism. This case may correspond to the state of FSRQs, with an Eddington ratio being higher than $0.1$ (Ghisellini et al. 2010; Xiong \& Zhang 2014). Our analysis results well support this framework.

Note that the above discussion is based on the results that the BH masses in GeV NLS1 galaxies are comparable to that in blazars. It has been widely suggested that the accretion of NLS1 galaxies may be close to the Eddington rate, or even super-Eddington rate (Boroson 2002; Collin \& Kawaguchi 2004; Xu et al. 2012). However, the ``super-Eddington" issue of RL NLS1 galaxies may be due to the underestimates of the BH masses and the overestimates of the accretion disk luminosities (Calderone et al. 2013). The bolometric luminosity of RL sources is contaminated by the jet emission, and it is not a good proxy of the disk luminosity. As we show in this analysis, the derived Eddington ratios of the GeV NLS1 galaxies are closed to FSRQs.

\acknowledgments
We thank the anonymous referee for his/her valuable suggestions. We appreciate helpful discussion with Bi-Fang Liu, Er-Lin Qiao, Xin-Wu Cao, Feng Yuan, and Yi Xie. This work is supported by the National Basic Research Program (973 Programme) of China (grant 2014CB845800), the National Natural Science Foundation of China (grants 11025313, 11373036, 11133002, and 11403005), the Strategic Priority Research Program "The Emergence of Cosmological Structures" of the Chinese Academy of Sciences (grant XDB09000000), the Guangxi Science Foundation (2013GXNSFFA019001 and 2014GXNSFBA118004), the Key Laboratory for the Structure and Evolution of Celestial Objects of Chinese Academy of Sciences, and the Young Researcher Grant of National Astronomical Observatories, Chinese Academic of Science. Shuang-Nan Zhang acknowledges support from the Qianren start-up grant 292012312D1117210.

\clearpage

\begin{deluxetable}{lccccccccccc}
\tablecaption{Derived Parameters from Our SED Fits with the Single-zone Leptonic Model}
\tablewidth{0pt}
\tabletypesize{\tiny}
\tablehead{
\colhead{Source\tablenotemark{\rm a}} &
\colhead{$p_{1}$} &
\colhead{$p_{2}$} &
\colhead{$\gamma_{\rm min}$} &
\colhead{$\gamma_{\rm max}$} &
\colhead{$\gamma_{\rm b}$} &
\colhead{$N_{0}$} &
\colhead{$\delta$} &
\colhead{$B$ [G]}&
\colhead{$R_{\rm in}\tablenotemark{\rm b}$}&
\colhead{$\log M_{\rm BH}/M_{\rm\odot}$}&
\colhead{$L_{\rm d}/L_{\rm Edd}$\tablenotemark{\rm c}}}

\startdata
1H 0323+342 (1)&-1.0&4.4&170$\pm$35&4000&883$\pm$361&(1.3$^{+2.6}_{-1.3}$)E-3&2.8$\pm$0.6&3.7$\pm$1.4&3.0&8.6&0.02\\
1H 0323+342 (2)&-1.0&4.6&190$\pm$40&5000&591$\pm$273&(2.4$^{+6.1}_{-2.4}$)E-3&3.6$\pm$1.3&2.1$\pm$1.2&3.0&8.6&0.01\\
1H 0323+342 (3)&-1.0&4.2&180$\pm$33&4000&383$\pm$160&(1.9$^{+3.6}_{-1.9}$)E-3&4.9$\pm$0.8&2.5$\pm$0.9&3.0&8.6&0.01\\
1H 0323+342 (4)&-1.0&4.2&180$\pm$25&4000&378$\pm$151&(5.4$^{+8.9}_{-5.4}$)E-3&4.5$\pm$0.6&1.9$\pm$0.6&3.0&8.6&0.02\\
1H 0323+342 (5)&-1.0&4.0&40$\pm$15&5000&271$\pm$60&(2.7$\pm$2.6)E-3&6.2$\pm$0.6&2.5$\pm$0.7&3.0&8.6&0.01\\
PMN J0948+0022 (1)&1.6&4.8&23$\pm$15&1700&267$\pm$109&(3.6$^{+4.5}_{-3.6}$)E-3&11.0$\pm$1.4&4.0$\pm$1.5&0.55&8.91&0.09\\
PMN J0948+0022 (2)&1.6&4.6&16$\pm$9&1900&260$\pm$80&(4.0$^{+4.2}_{-4.0}$)E-3&10.8$\pm$1.3&5.8$\pm$1.4&0.55&8.91&0.16\\
PMN J0948+0022 (3)&1.6&4.0&43$\pm$13&2000&336$\pm$126&(9.1$^{+12.0}_{-9.1}$)E-3&8.6$\pm$1.3&4.6$\pm$1.2&0.55&8.91&0.16\\
PMN J0948+0022 (4)&1.6&4.8&47$\pm$9&1700&202$\pm$69&(5.1$\pm$5.0)E3&11.1$\pm$1.0&5.1$\pm$1.2&0.55&8.91&0.16\\
PMN J0948+0022 (5)&1.6&4.6&43$\pm$15&1700&234$\pm$64&(4.9$\pm$3.6)E3&11.6$\pm$0.8&3.7$\pm$0.7&0.55&8.91&0.16\\
PMN J0948+0022 (6)&1.2&3.8&43$\pm$13&3000&285$\pm$59&(1.1$\pm$0.7)E3&9.5$\pm$0.5&2.4$\pm$0.4&0.55&8.91&0.16\\
PMN J0948+0022 (7)&1.2&4.4&108$\pm$18&4000&186$\pm$63&(1.7$^{+1.8}_{-1.7}$)E-3&13.5$\pm$1.1&1.7$\pm$0.8&0.55&8.91&0.16\\
PMN J0948+0022 (8)&1.2&4.4&80$\pm$19&4000&145$\pm$63&(2.0$^{+2.9}_{-2.0}$)E-3&13.7$\pm$1.8&2.1$\pm$1.1&0.55&8.91&0.09\\
PMN J0948+0022 (9)&1.6&4.2&20$\pm$15&2000&350$\pm$147&(3.2$^{+5.3}_{-3.2}$)E-3&11.4$\pm$2.2&4.2$\pm$2.0&0.55&8.91&0.16\\
SBS 0846+513&1.2&3.5&66$\pm$28&5500&366$\pm$84&(7.6$\pm$6.5)E2&7.4$\pm$0.8&2.0$\pm$0.6&\nodata&7.72&$<0.05$\\
PKS 1502+036&2.0&4.2&40$\pm$18&2000&277$\pm$60&(5.8$\pm$4.2)E3&9.5$\pm$0.8&4.7$\pm$0.3&1.0&8.3&0.03\\
PKS 2004$-$447&1.4&4.2&36$\pm$10&3000&359$\pm$69&(6.0$\pm$4.0)E2&6.4$\pm$0.5&4.1$\pm$0.6&\nodata&6.73&$<0.05$\\
\enddata
\tablenotetext{\rm a}{The numbers from (1) to (5) for 1H 0323+342 mark the SEDs observed in MJD (54682-55048), MJD (54775-54805), MJD (56262-56310), MJD (56470-56500), and MJD (56531-56535), respectively. The numbers from (1) to (9) for PMN J0948+0022 mark the SEDs observed in MJD 54805, MJD 54916, MJD 54956, MJD 54976 (or MJD 54986), MJD 54996 (or MJD 54936), MJD 54916-55015, MJD 55385, MJD 55733, and MJD 55843-55846, respectively.}
\tablenotetext{\rm b}{$R_{\rm in}$ is in unit of the Schwarzschild radius.}
\tablenotetext{\rm c} {The Eddington ratio of 1H 0323+342 is calculated with $(L_{\rm d}+L_{\rm corona})/L_{\rm Edd}$.}
\end{deluxetable}

\begin{deluxetable}{lccccccc}
\tabletypesize{\scriptsize}
\tablecaption{Jet Powers Carried by Different Ingredients and the Derived Radiation Efficiency and Magnetization Parameter}
\tablewidth{0pt}
\tabletypesize{\tiny}
\tablehead{
\colhead{Source\tablenotemark{\rm a}} &
\colhead{$\log P_{\rm e}$} &
\colhead{$\log P_{\rm p}$} &
\colhead{$\log P_{B}$} &
\colhead{$\log P_{\rm r}$} &
\colhead{$\log L_{\rm d}$} &
\colhead{$\varepsilon$} &
\colhead{$\sigma$}\\
\colhead{}& \colhead{(erg s$^{-1}$)} & \colhead{(erg s$^{-1}$)} & \colhead{(erg s$^{-1}$)}& \colhead{(erg s$^{-1}$)} & \colhead{(erg s$^{-1}$)} & \colhead{} & \colhead{}}

\startdata
1H 0323+342 (1)&43.32$\pm$0.49&43.66$\pm$0.51&43.26$\pm$0.53&44.08$\pm$0.19&44.66&0.58$\pm$0.35&0.10$^{+0.12}_{-0.10}$\\
1H 0323+342 (2)&43.50$\pm$0.71&44.00$\pm$0.65&43.18$\pm$0.81&43.55$\pm$0.33&44.66&0.20$^{+0.23}_{-0.20}$&0.09$^{+0.17}_{-0.09}$\\
1H 0323+342 (3)&43.35$\pm$0.49&44.00$\pm$0.48&43.88$\pm$0.46&43.51$\pm$0.16&44.66&0.14$\pm$0.10&0.49$^{+0.52}_{-0.49}$\\
1H 0323+342 (4)&43.67$\pm$0.45&44.32$\pm$0.45&43.49$\pm$0.39&43.71$\pm$0.13&44.66&0.15$\pm$0.11&0.10$\pm$0.09\\
1H 0323+342 (5)&43.50$\pm$0.28&44.33$\pm$0.27&44.31$\pm$0.30&43.82$\pm$0.09&44.66&0.13$\pm$0.06&0.66$\pm$0.45\\
PMN J0948+0022 (1)&44.29$\pm$0.41&45.70$\pm$0.49&45.38$\pm$0.38&44.70$\pm$0.12&45.97&0.06$\pm$0.05&0.42$\pm$0.36\\
PMN J0948+0022 (2)&44.35$\pm$0.33&45.84$\pm$0.41&45.65$\pm$0.29&44.71$\pm$0.12&46.20&0.04$\pm$0.03&0.58$\pm$0.39\\
PMN J0948+0022 (3)&44.31$\pm$0.37&45.43$\pm$0.40&45.07$\pm$0.34&44.69$\pm$0.14&46.20&0.11$\pm$0.07&0.35$\pm$0.27\\
PMN J0948+0022 (4)&44.27$\pm$0.32&45.49$\pm$0.34&45.6$\pm$0.26&44.62$\pm$0.10&46.20&0.05$\pm$0.03&1.08$\pm$0.65\\
PMN J0948+0022 (5)&44.41$\pm$0.25&45.63$\pm$0.29&45.41$\pm$0.21&44.82$\pm$0.07&46.20&0.08$\pm$0.04&0.50$\pm$0.24\\
PMN J0948+0022 (6)&44.41$\pm$0.21&45.47$\pm$0.23&44.69$\pm$0.17&44.74$\pm$0.06&46.20&0.13$\pm$0.05&0.13$\pm$0.05\\
PMN J0948+0022 (7)&44.78$\pm$0.47&45.78$\pm$0.44&45.00$\pm$0.39&45.38$\pm$0.12&46.20&0.24$\pm$0.16&0.11$\pm$0.10\\
PMN J0948+0022 (8)&44.74$\pm$0.54&45.87$\pm$0.53&45.22$\pm$0.47&45.20$\pm$0.15&45.97&0.14$\pm$0.13&0.17$\pm$0.19\\
PMN J0948+0022 (9)&44.39$\pm$0.50&45.76$\pm$0.56&45.48$\pm$0.52&44.97$\pm$0.21&46.20&0.09$\pm$0.09&0.43$^{+0.52}_{-0.43}$\\
SBS 0846+513&44.02$\pm$0.27&44.93$\pm$0.28&44.20$\pm$0.32&44.51$\pm$0.10&$<44.51$&0.22$\pm$0.10&0.12$\pm$0.09\\
PKS 1502+036&43.47$\pm$0.22&44.70$\pm$0.30&45.36$\pm$0.16&44.04$\pm$0.08&44.84&0.04$\pm$0.01&3.58$\pm$1.32\\
PKS 2004$-$447&43.29$\pm$0.18&44.38$\pm$0.21&44.66$\pm$0.19&43.61$\pm$0.07&$<43.57$&0.05$\pm$0.02&1.53$\pm$0.67\\
\enddata
\tablenotetext{\rm a}{Same as table 1.}

\end{deluxetable}

\clearpage

\begin{figure*}
\includegraphics[angle=0,scale=0.55]{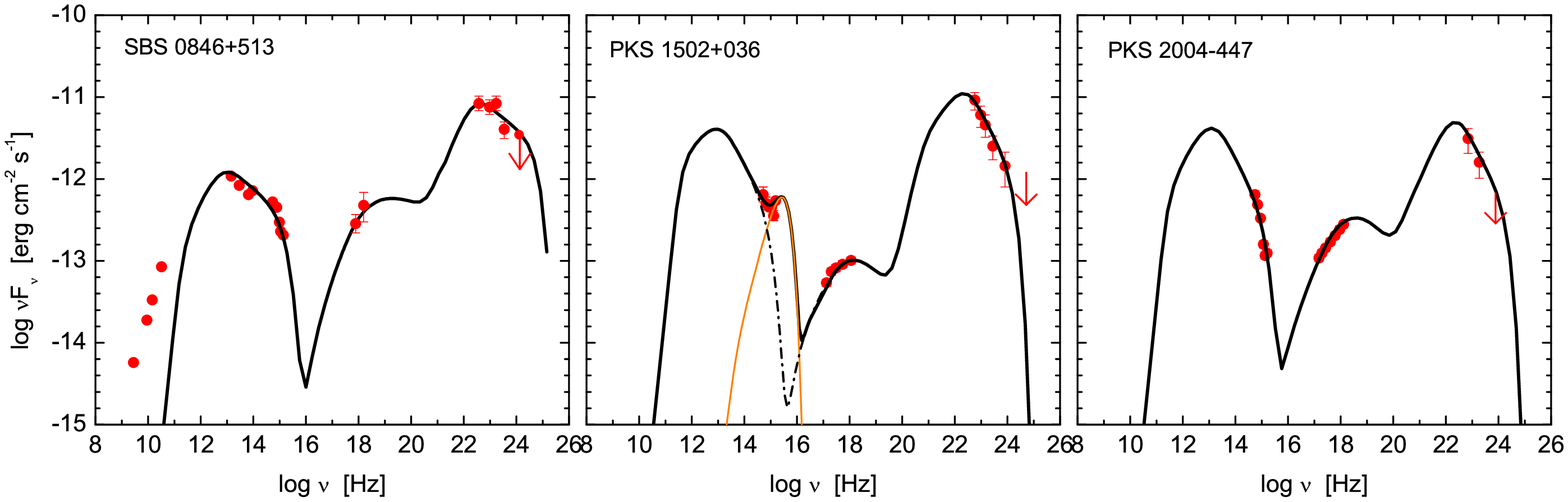}

\bf{Fig. 1.---(b) Observed SEDs (red data points) with the model best-fits (lines) for SBS 0846+513, PKS 1502+036, and PKS 2004$-$447, respectively.
The styles of the lines are the same as Fig. 1 (a).}
\end{figure*}

\begin{figure*}
\includegraphics[angle=0,scale=0.55]{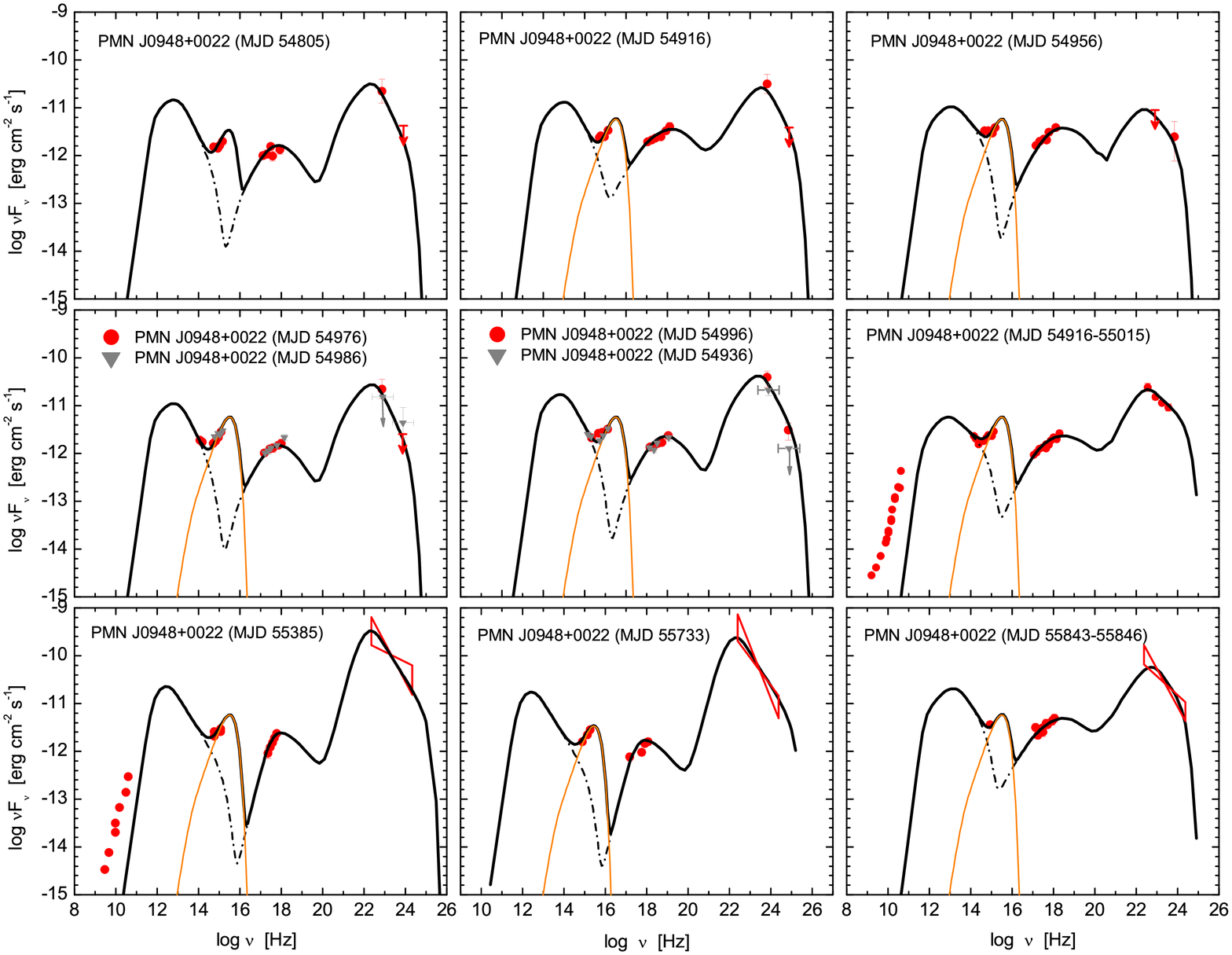}

\bf{Fig. 1.---(c) Observed SEDs (red data points and bow-ties) with the model best-fits (lines) for PMN J0948+0022 observed in nine campaigns.
The styles of the lines are the same as Fig. 1 (a). The SEDs observed in the campaigns MJD 54986 and MJD 54936 are consistent with
the SEDs obtained in the campaigns MJD 54976 and MJD 54996 within the error bars, respectively. Their data are also shown with grey triangles,
but they are not included in our SED fits.}
\end{figure*}

\begin{figure*}\centering
\includegraphics[angle=0,scale=0.3]{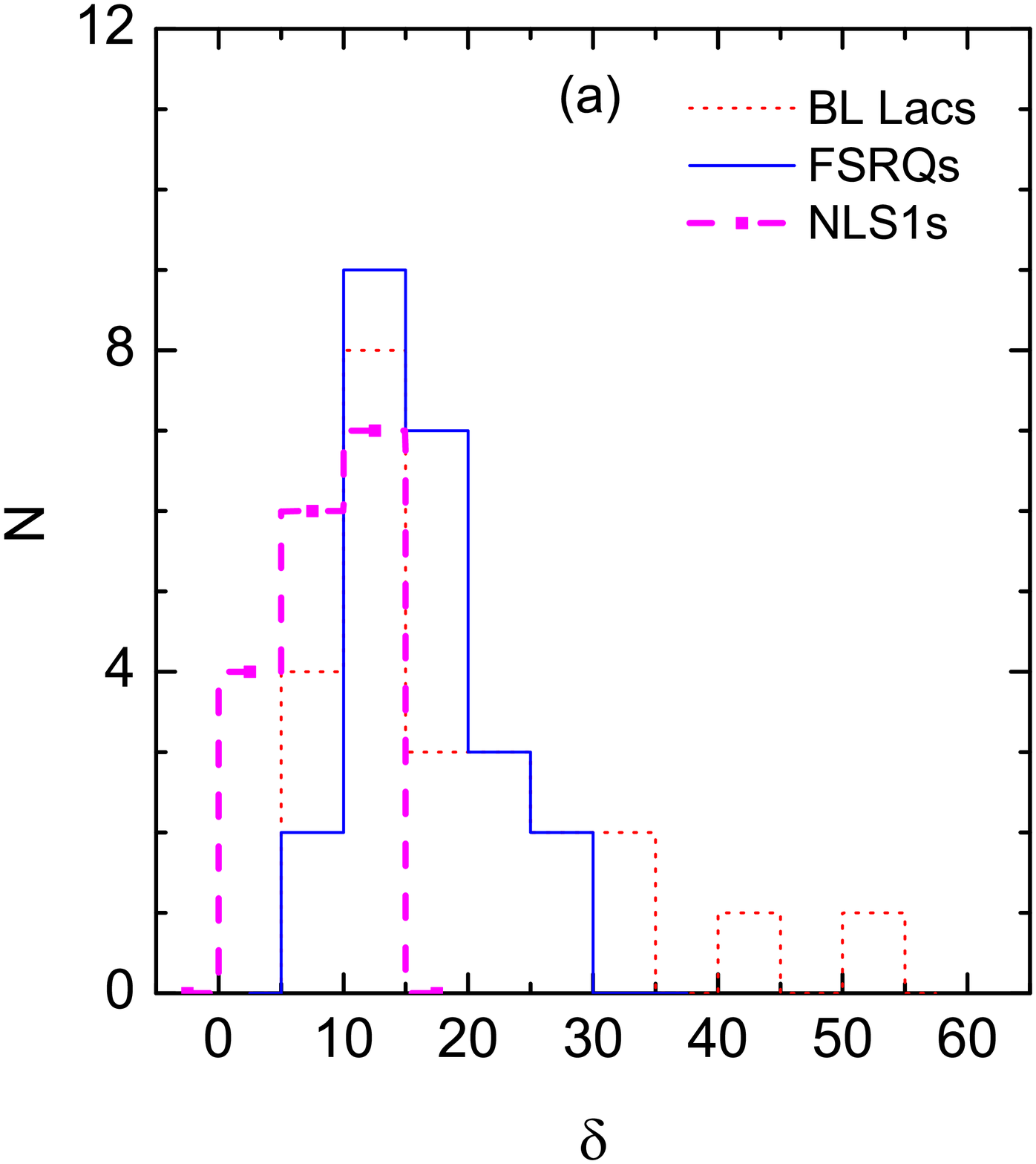}
\includegraphics[angle=0,scale=0.3]{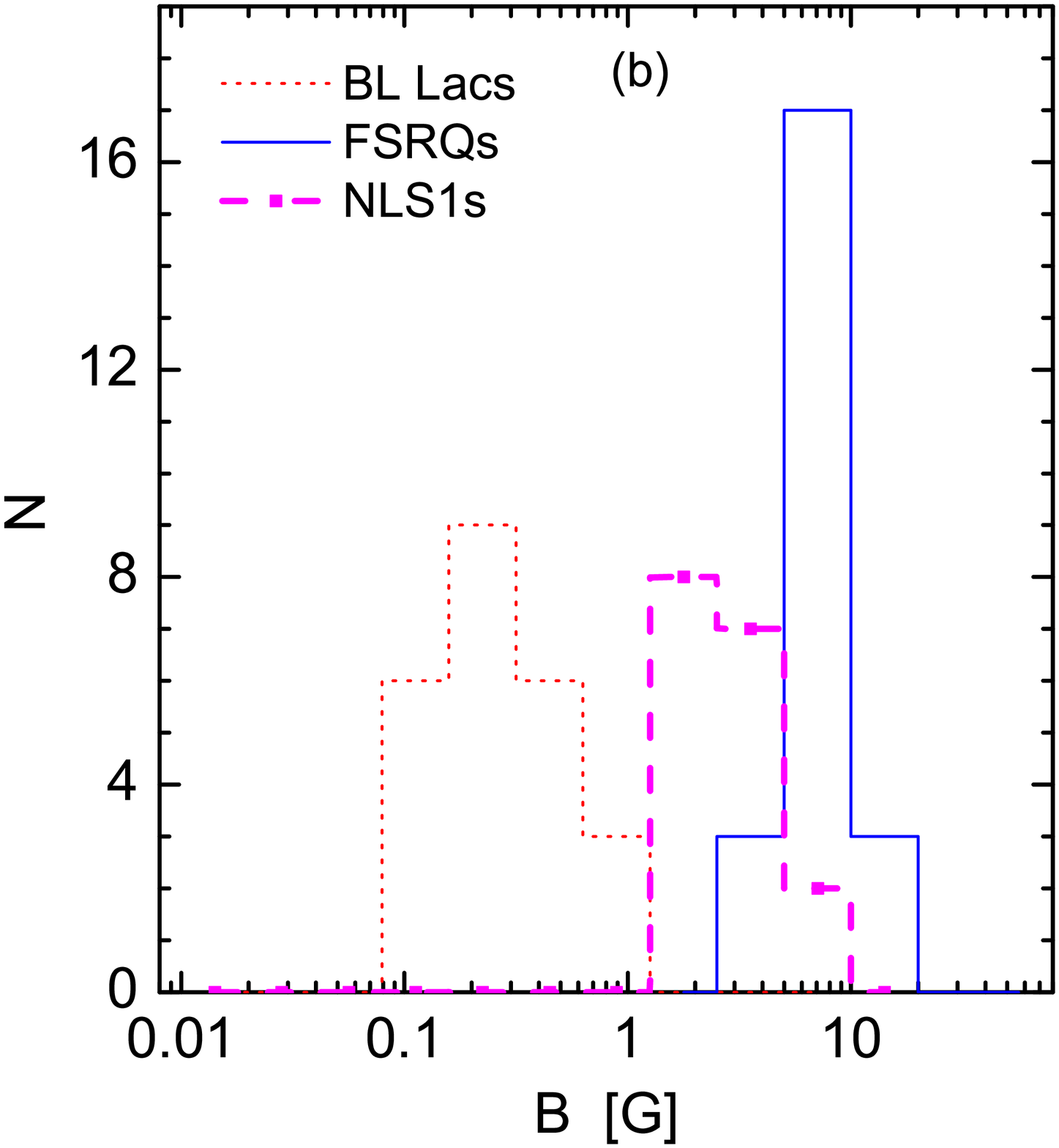}\\
\includegraphics[angle=0,scale=0.3]{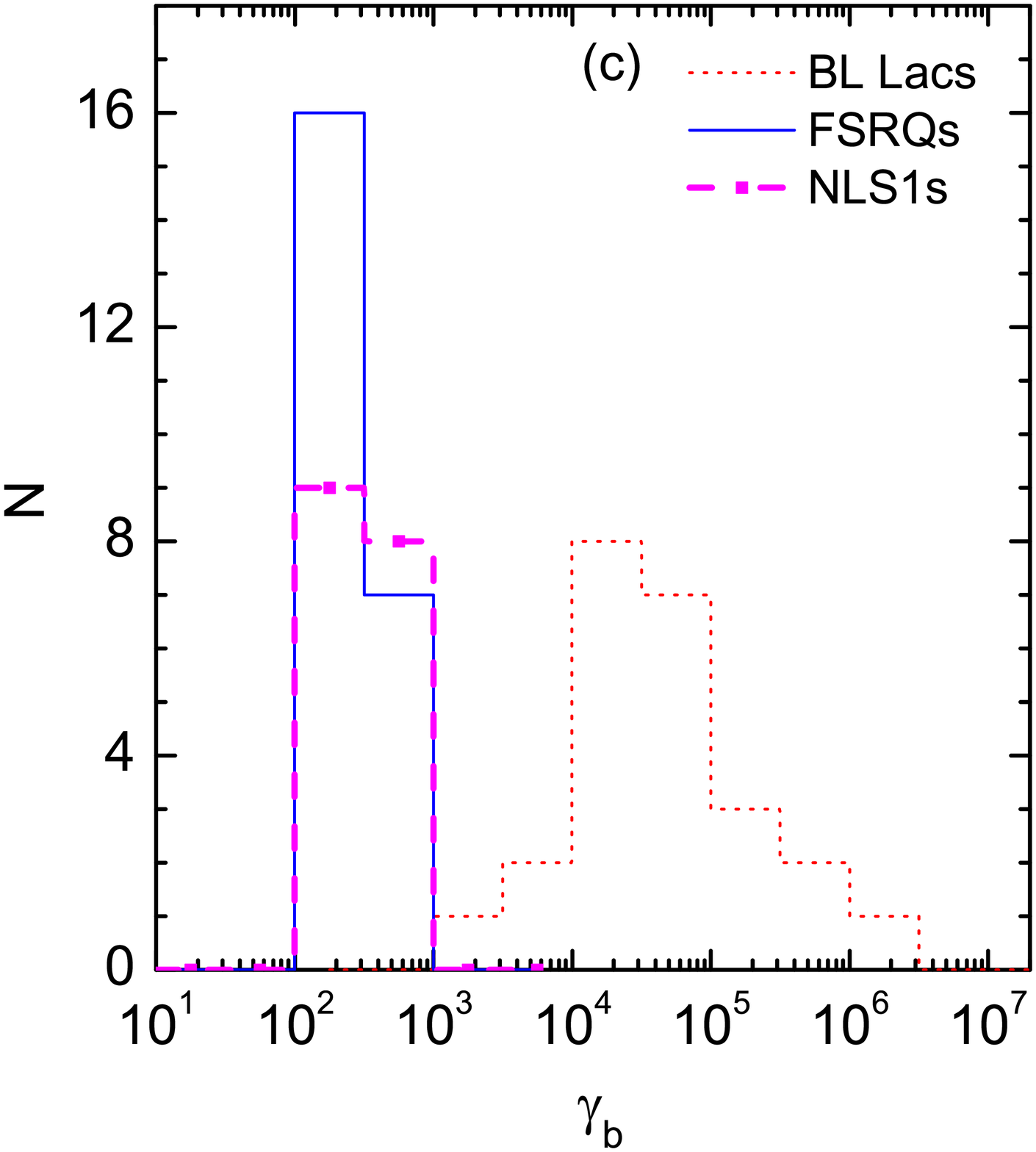}
\includegraphics[angle=0,scale=0.3]{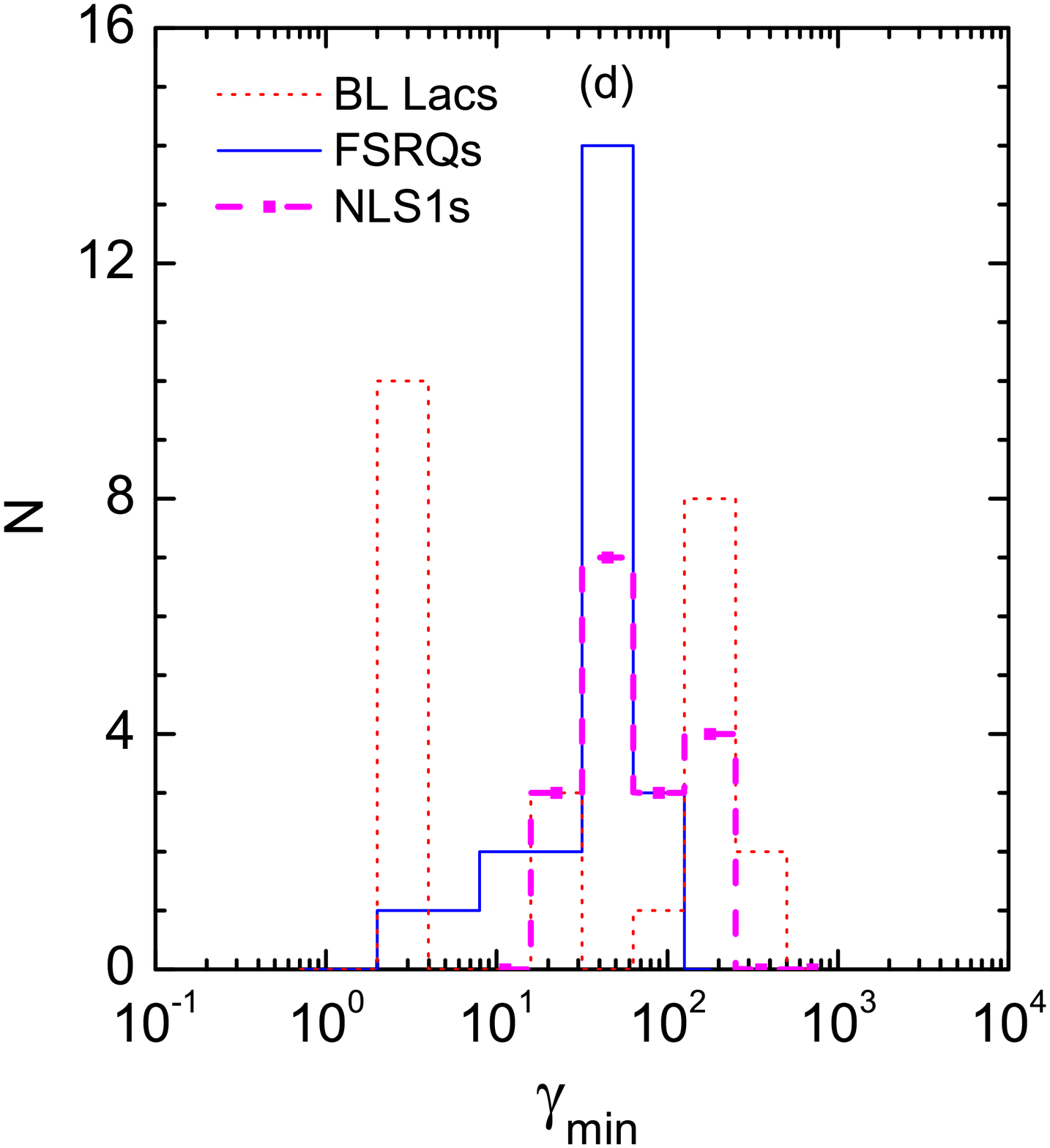}
\caption{Distributions of the beaming factor (a), the magnetic field strength (b), the break Lorentz factor of electrons (c), and the minimum Lorentz factor of electrons (d). The blazar data are taken from Zhang et al. (2014).}\label{jet-parameters}
\end{figure*}

\begin{figure*}\centering
\includegraphics[angle=0,scale=0.35]{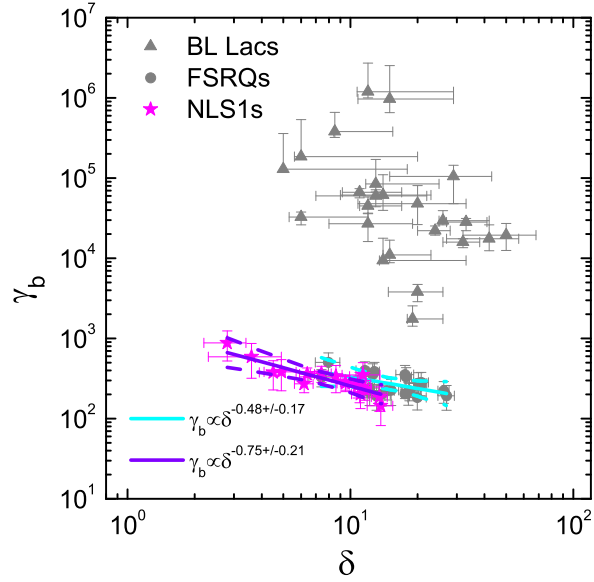}
\caption{$\gamma_{\rm b}$ as a function of $\delta$. The solid lines are the best-fits
to the data of FSRQs (cyan line) and NLS1 galaxies (violet line), respectively, and the dashed lines mark the corresponding 3$\sigma$ confidence bands of the best-fits. The blazar data are taken from Zhang et al. (2014).}\label{gammab-delta}
\end{figure*}

\begin{figure*}
\includegraphics[angle=0,scale=0.53]{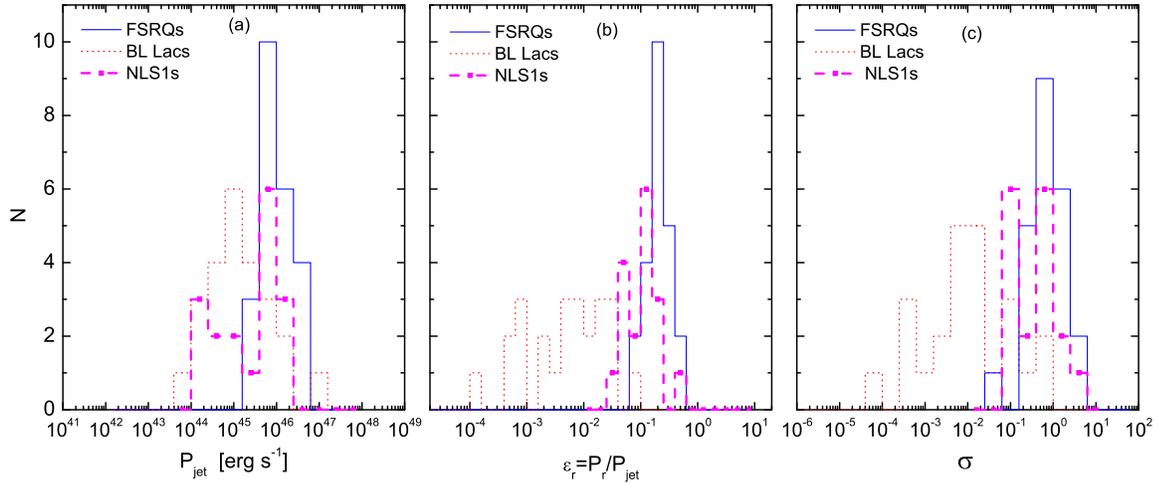}
\caption{Distributions of the jet powers (a), the jet radiation efficiencies (b), and the magnetization parameters (c) of GeV NLS1 galaxies in comparison with those of the BL Lacs and FSRQs. The blazar data are taken from Zhang et al. (2014).}\label{Pjet}
\end{figure*}

\begin{figure*}
\includegraphics[angle=0,scale=0.5]{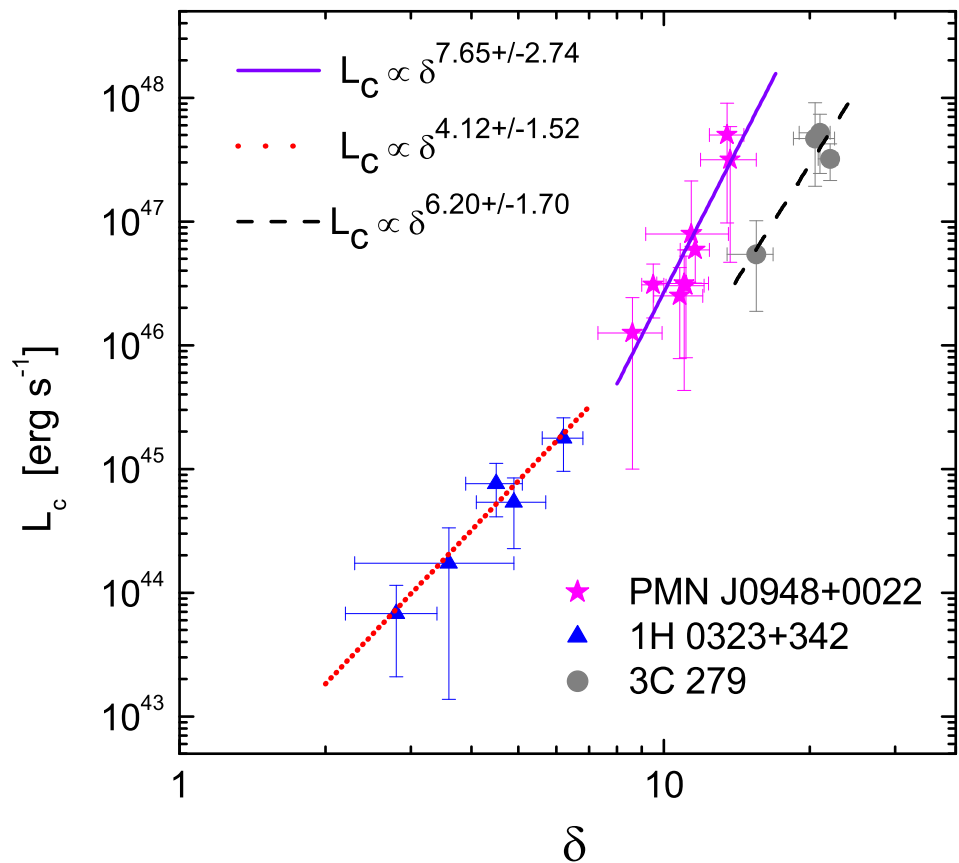}
\caption{EC peak luminosity in the observer's frame as a function of the beaming factor for PMN J0948+0022 and 1H 0323+342. The violet solid and red dotted lines are the best-fits to the data by considering the errors in both X and Y axis. The data of 3C 279 with the best-fit line (black dashed line) are also shown for comparison. Its data are from Zhang et al. (2013b).}\label{Lc-delta}
\end{figure*}

\begin{figure*}
\includegraphics[angle=0,scale=0.3]{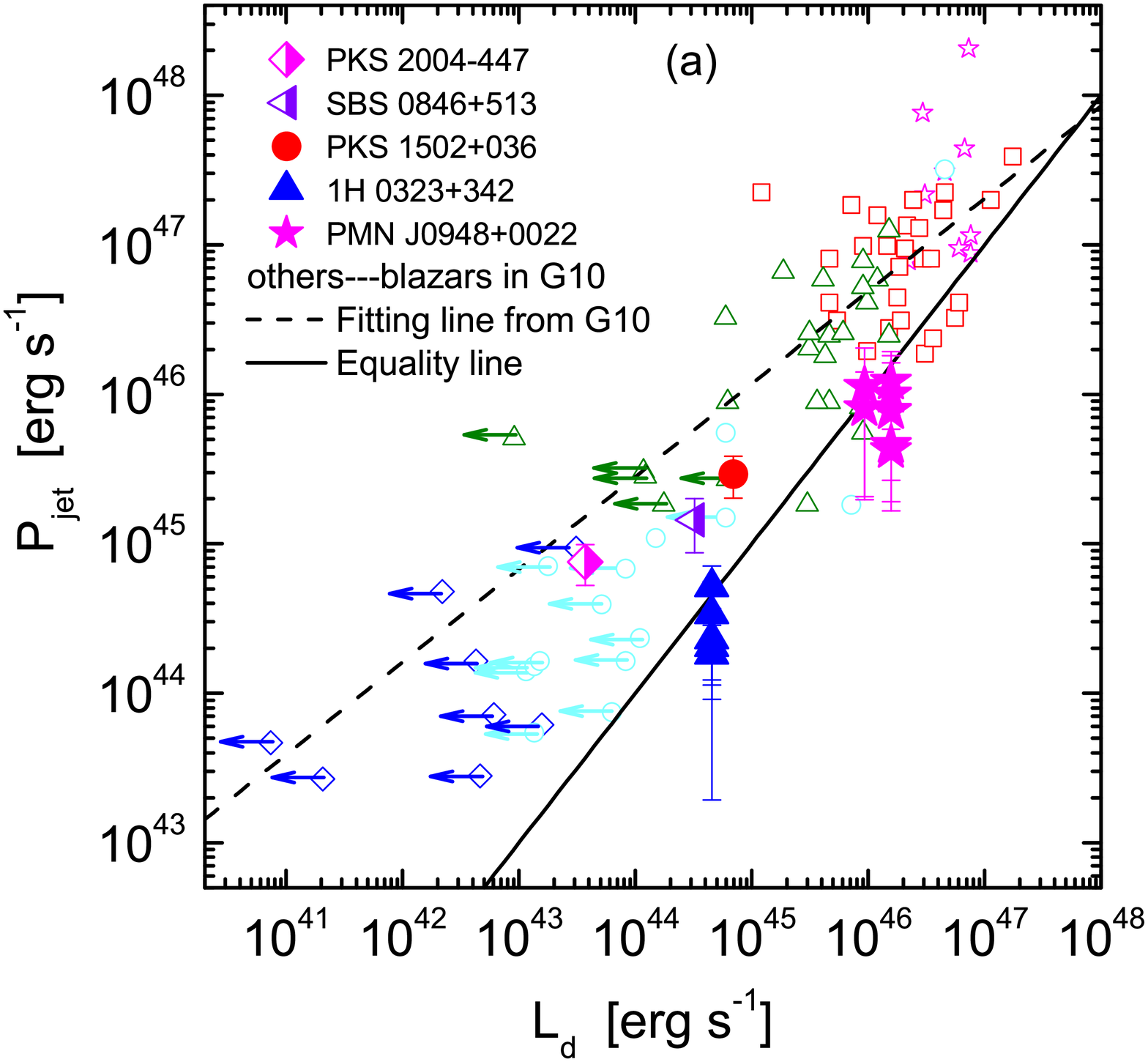}
\includegraphics[angle=0,scale=0.3]{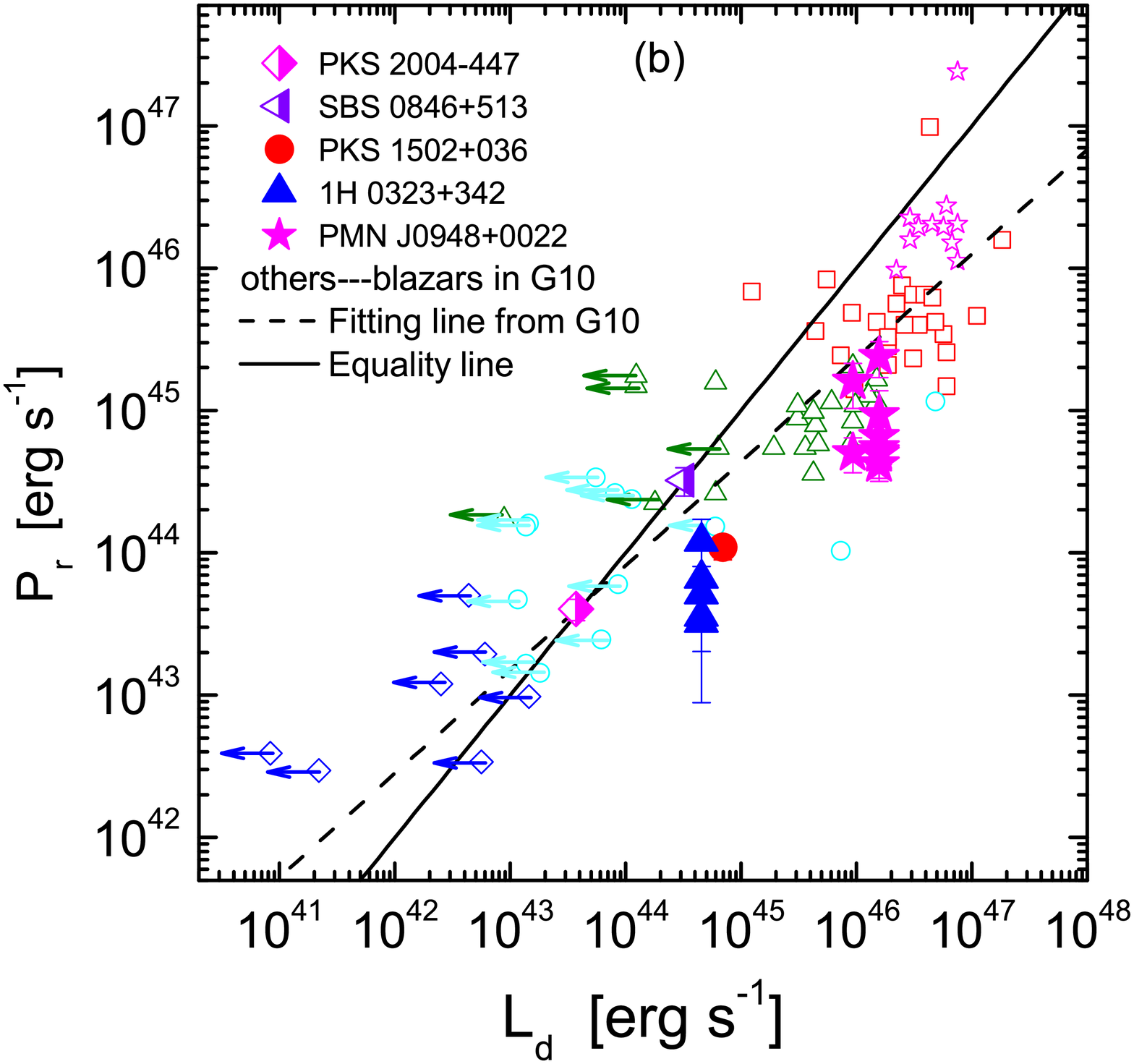}
\caption{Jet total power (a) and jet radiation power (b) as a function of the accretion disk luminosity. The solid lines are the equality lines. The blazar data (opened symbols) and the fitting lines (dashed lines) for FSRQs are taken from Ghisellini et al. (2010). \textbf{The opened symbols with different colors correspond to blazars with different $\gamma$-ray observed luminosities: stars for $L_{\gamma}>10^{48.5}$, squares for $10^{47.5}<L_{\gamma}<10^{48.5}$, triangles for $10^{46.5}<L_{\gamma}<10^{47.5}$, circles for $10^{45.5}<L_{\gamma}<10^{46.5}$, and diamonds for $L_{\gamma}<10^{45.5}$.}}\label{Pjet-Ld}
\end{figure*}

\begin{figure*}
\includegraphics[angle=0,scale=0.3]{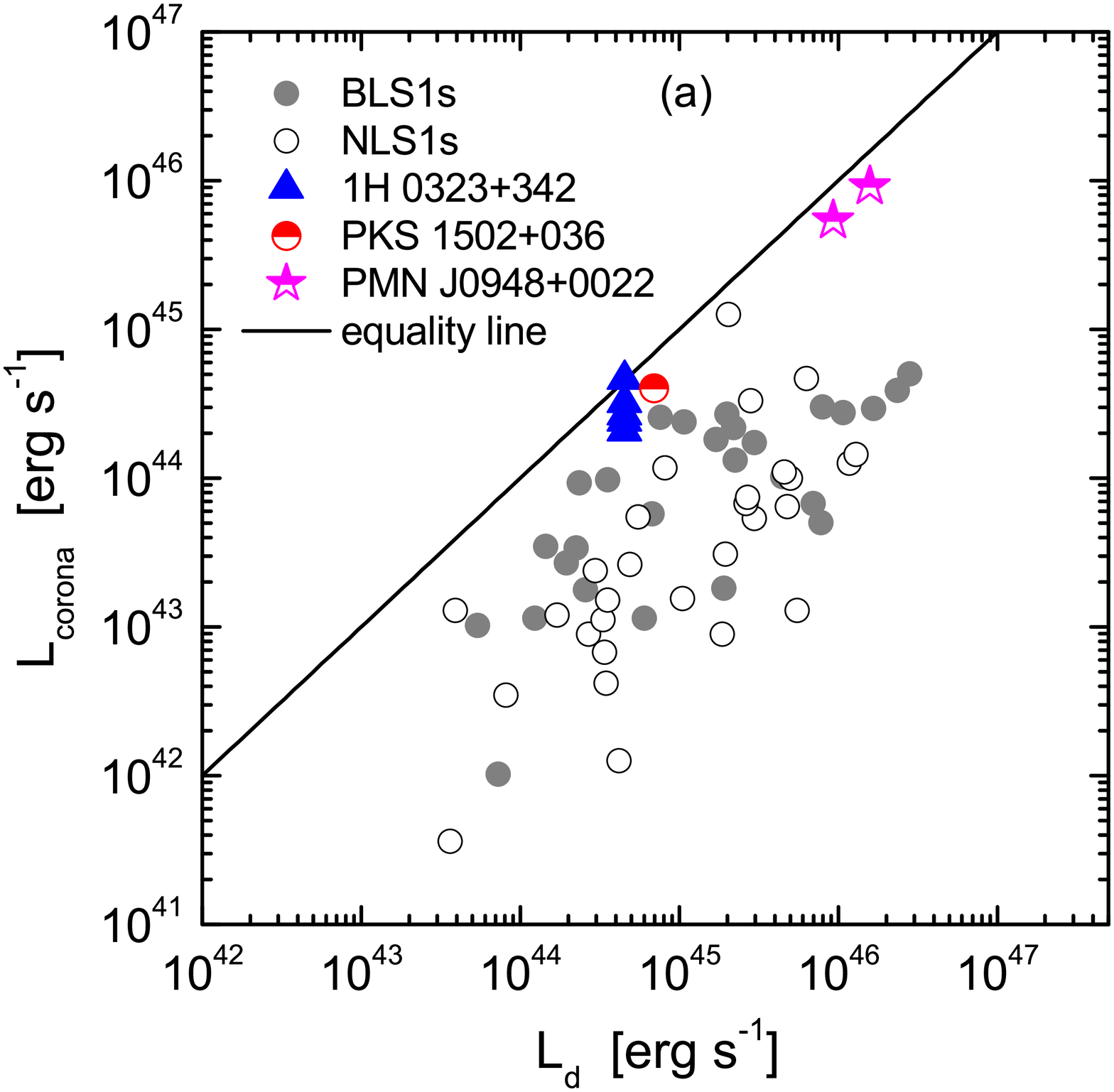}
\includegraphics[angle=0,scale=0.3]{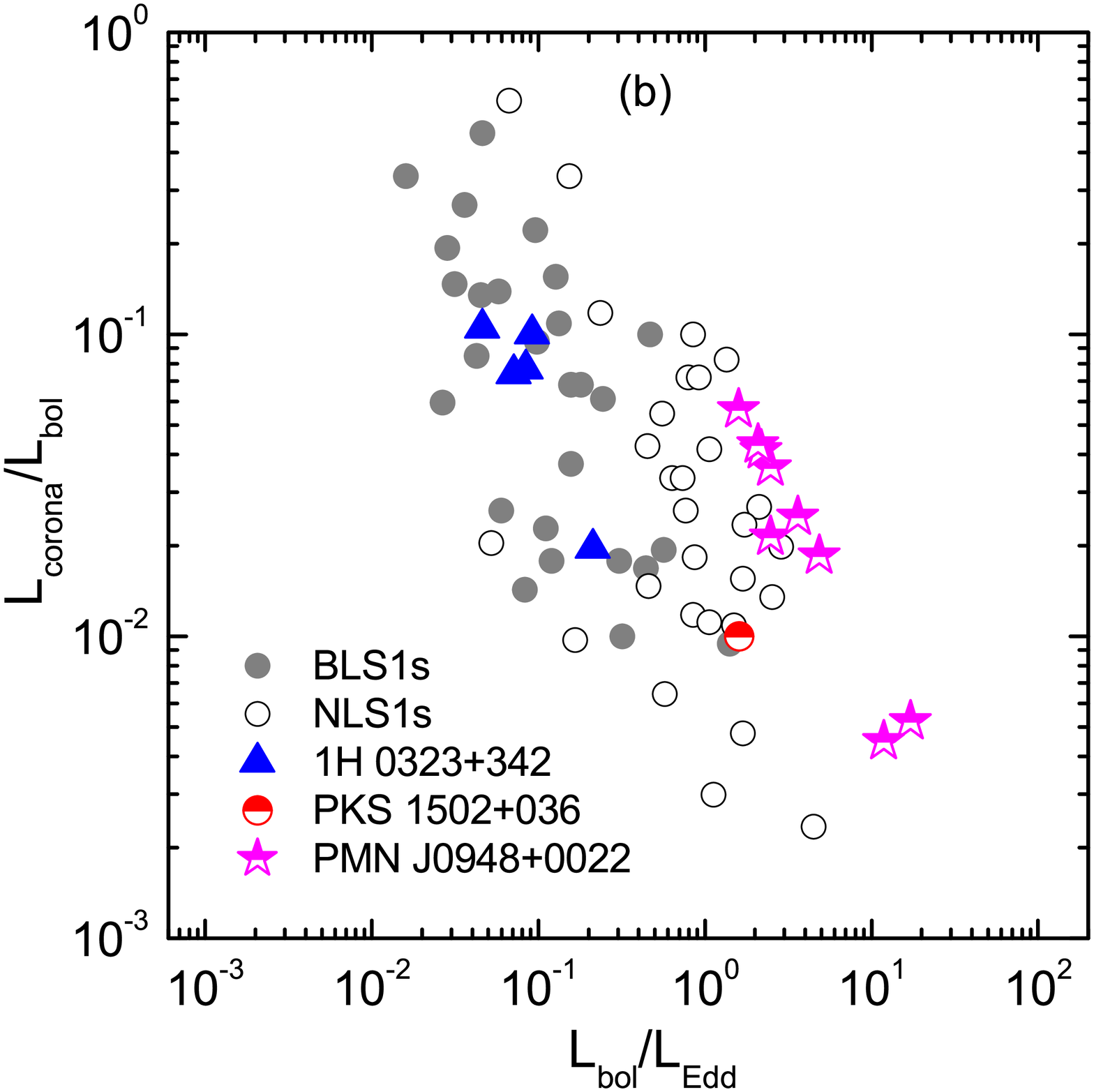}
\caption{(a) Corona luminosity as a function of the accretion disk luminosity. (b) Ratio of the corona luminosity to the bolometric luminosity against the ratio of the bolometric luminosity to the Eddington luminosity. For PKS 1502+036 and PMN J0948+0022, only upper limits of corona luminosity ($L_{\rm corona}=0.58*L_{\rm d}$) are present, since no significant corona emission is observed in their SEDs. The data of RQ BLS1 and NLS1 galaxies are taken from Wang et al. (2004).}\label{Lcorona}
\end{figure*}

\begin{figure*}
\includegraphics[angle=0,scale=0.3]{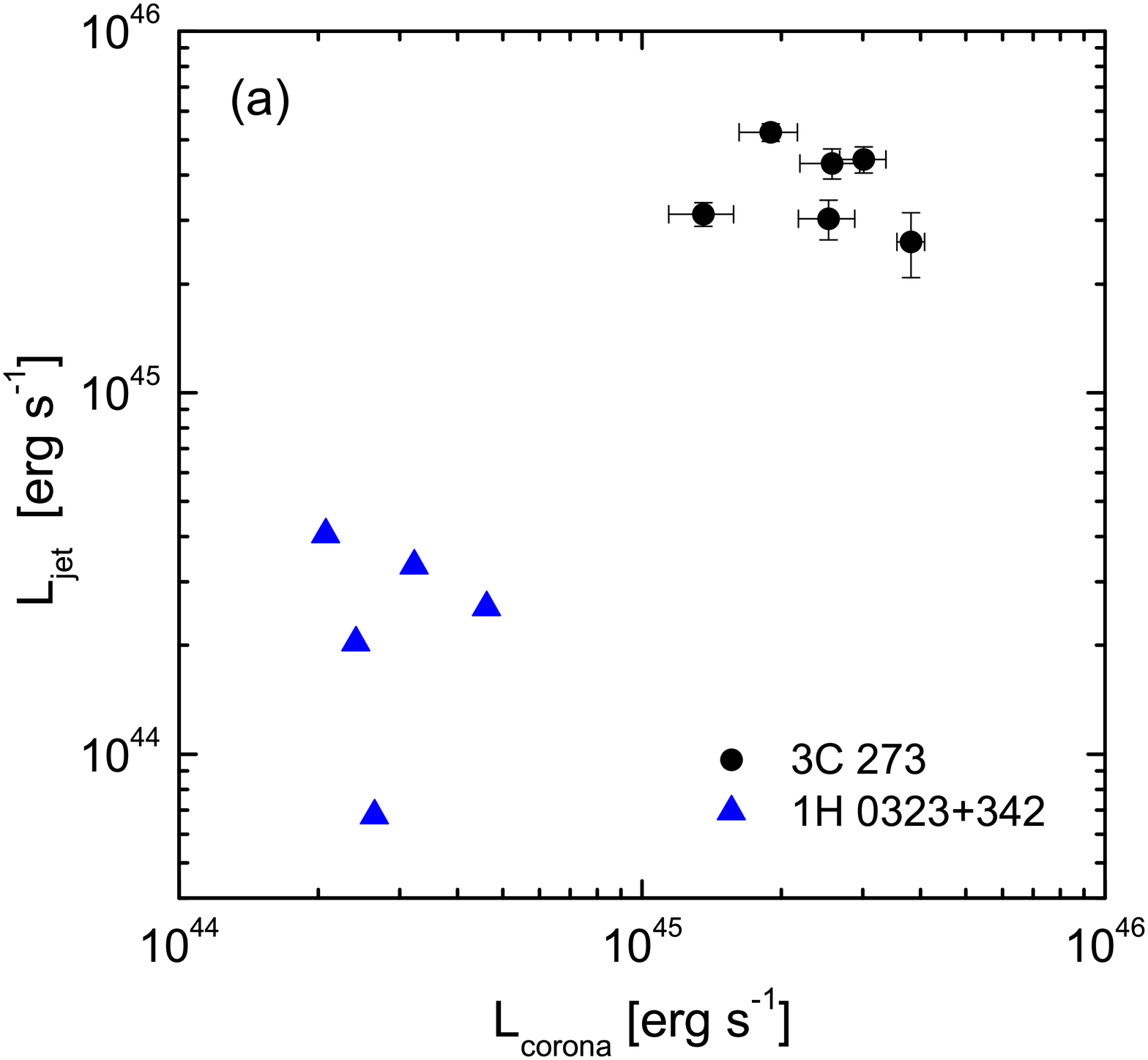}
\includegraphics[angle=0,scale=0.3]{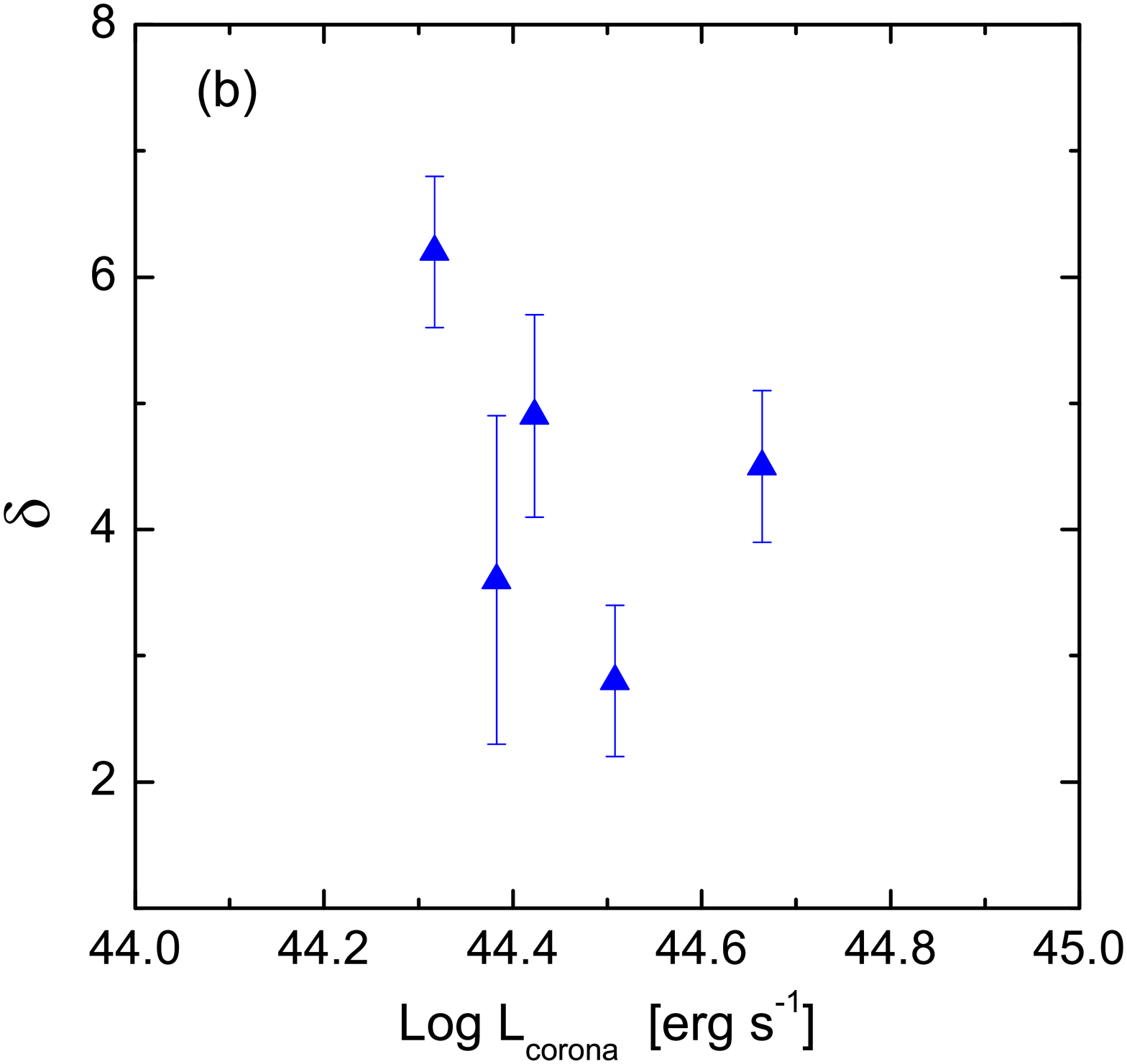}
\caption{(a) Jet luminosity ($L_{\rm jet}$) as a function of the corona luminosity ($L_{\rm corona}$) for 1H 0323+342 and 3C 273. The jet luminosities are taken from the same energy band of corona emission. $L_{\rm jet}$ and $L_{\rm corona}$ of 1H 0323+342 are taken from $10^{16}$ Hz to 150 keV. The data of 3C 273 are taken from Grandi \& Palumbo (2004) in energy band of 2--10 keV. (b) $\delta$ as a function of $L_{\rm corona}$ for 1H 0323+342.}\label{Ljet-Lcorona}
\end{figure*}

\end{document}